\documentclass{timgad}
\pdfoutput=1
\usepackage{tikz}
\usetikzlibrary{shapes,backgrounds,calc,arrows,fadings}
\usepackage{ifpdf} 

\title{Homomorphism reconfiguration via homotopy}
\date{First appeared: August 2014, updated: March 2017.}
\author{Marcin Wrochna}
\affil{Institute of Informatics, University of Warsaw, Poland\\
	\textnormal{\texttt{m.wrochna@mimuw.edu.pl}}}
\keywords{reconfiguration, recoloring, homomorphism, Hom-complex}

\usepackage[backend=bibtex,bibencoding=ascii,style=alphabetic,bibstyle=alphabetic,giveninits=true,doi=false,isbn=false,url=false,maxbibnames=99]{biblatex}
\renewbibmacro{in:}{}
\AtEveryBibitem{\clearname{editor}}

\newbibmacro{string+link}[1]{%
	\iffieldundef{ee}
	{\iffieldundef{url}
		{\iffieldundef{doi}
			{\iffieldundef{eprint}
				{#1}
				{\href{http://arxiv.org/abs/\thefield{eprint}}{#1}}}
			{\href{http://dx.doi.org/\thefield{doi}}{#1}}}
		{\href{\thefield{url}}{#1}}}
	{\href{\thefield{ee}}{#1}}} 
\DeclareFieldFormat{title}{\usebibmacro{string+link}{\mkbibemph{#1}}}
\DeclareFieldFormat[article,inproceedings,inbook,incollection,mastersthesis,thesis]{title}{\usebibmacro{string+link}{\mkbibquote{#1}}}

\newcommand{\Oh}{\ensuremath{\mathcal{O}}}
\newcommand{\Hom}{\ensuremath{\mbox{Hom}}}

\def\red#1{\overline{#1}}

\def\problem#1{\textsc{#1}}
\def\cclass#1{{#1}}

\include{figureDefs}

\begin{document}

\maketitle

\begin{abstract}
For a fixed graph $H$, we consider the \problem{$H$-Recoloring} problem :
given a graph $G$ and two $H$-colorings of $G$, i.e., homomorphisms from $G$ to $H$, can one be transformed into the other by changing one color at a time, maintaining an $H$-coloring throughout.
This is the same as finding a path in the $\Hom(G,H)$ complex.
For $H=K_k$ this is the problem of finding paths between $k$-colorings, which was recently shown to be in \cclass{P} for $k\leq 3$ and \cclass{PSPACE}-complete otherwise.
We generalize the positive side of this dichotomy by providing an algorithm that solves the problem in polynomial time for any $H$ with no $C_4$ subgraph.
This gives a large class of constraints for which finding solutions to the Constraint Satisfaction Problem is \cclass{NP}-complete, but finding paths in the solution space is in \cclass{P}.

The algorithm uses a characterization of possible reconfiguration sequences (paths in $\Hom(G,H)$), whose main part is a purely topological condition described in terms of the fundamental groupoid of $H$ seen as a topological space. 
\end{abstract}

\section{Introduction}
\paragraph*{Reconfiguration}
Reconfiguration is a framework in which we study how discrete structures, constrained in various ways, can be carefully transformed with small steps.
This is often best described by finding paths in a solution graph, whose vertices are all solutions to a combinatorial problem and whose edges define the steps between solutions one is allowed to make.

For example, in \problem{$k$-Recoloring}~\cite{BonsmaC09,CerecedaHJ11,johnson2014colouring,BonsmaMNR14}, one is given two proper $k$-colorings of a graph $G$ and the question is whether one can be transformed into the other by changing the color of one vertex at a time, maintaining a proper coloring throughout.
In other words, the solution graph has proper $k$-colorings as vertices (solutions) and edges (reconfigurations steps) between any two colorings that differ only at one vertex of $G$.
Another well studied example is \problem{Token Jumping}~\cite{HearnD05,KaminskiMM12,BonsmaKW14}, where the solutions are independent sets of some given size (seen as sets of tokens on the graph's vertices) and a reconfiguration step removes one vertex from the set to add another (jumps one token to a different vertex).
Yet another example is the reconfiguration of generalized SAT problems~\cite{GopalanKMP09,Schwerdtfeger14,Schwerdtfeger16,MouawadNPR15}, where solutions are satisfying assignments of a given formula, and a reconfiguration step flips one variable of the assignment.

\paragraph*{Homomorphisms}
A homomorphism from a graph $G$ to a graph $H$ is a mapping $\sigma:V(G)\to V(H)$ such that edges are mapped to edges, that is, $uv\in E(G)$ implies $\sigma(u)\sigma(v)\in E(H)$.
We also use the name $H$-coloring, especially when $H$ is fixed in the context. Vertices of $H$ are then called \emph{colors}.
Note that a $K_k$-coloring (where $H$ is the graph with all edges except loops) is the same as a $k$-coloring.

Define the solution graph $\Hom_1(G,H)$ to be the graph with $H$-colorings of $G$ as vertices and edges between any two $H$-colorings that differ in the color of only one vertex
(the name comes from it being a 1-dimensional restriction of a richer object, called the $\Hom(G,H)$ complex in combinatorial algebraic topology, see below).
For a fixed graph $H$, \problem{$H$-Recoloring} is the problem asking whether two given $H$-colorings of a given graph $G$ are connected by a path in $\Hom_1(G,H)$.
\problem{Shortest $H$-Recoloring} asks whether there is such a path of at most some given length.


We use graph homomorphisms as a tool to explore how different constraints influence the complexity of reconfiguration.
As our aim is to give more general statements about reconfiguration,
they should be seen as a special case of Constraint Satisfaction Problems (CSPs), which can express a range of problems including \problem{$k$-Coloring}, generalized \problem{SAT} or \problem{Independent Set} (in weighted variants).
However, graph homomorphisms already display many features of general CSPs and arise naturally in various situations.
See \cite{nevsetril2007homomorphisms} for an excellent survey, or Hell and Ne{\v{s}}et{\v{r}}il's book~\cite{hell2004homomorphism_book} for an in-depth look into the subject.

This approach allowed the author to argue in~\cite{Wrochna14} that the only notion of sparseness that can be applied algorithmically to (unparameterized) reconfiguration problems in general is treedepth, and that many such problems are PSPACE-complete even in graphs of bounded bandwidth (thus pathwidth, treewidth, etc.).
The reduction showed there explains why reconfiguration variants of easy combinatorial problems can be hard. This paper grew out of an attempt to find fundamental reasons for which reconfiguration variants of hard problems can be, quite surprisingly, easy.

\begin{figure}[H]
	\centering
	\begin{tikzpicture}[scale=0.6]
	\begin{scope}[shift={(-4,0)}]
		\node[G] (w5) at (0:1) {};	
		\node[G] (w1) at (72:1) {};
		\node[G] (w2) at (144:1) {};
		\node[G1] (w3) at (-144:1) {};
		\node[G] (w4) at (-72:1) {};						
		\draw (w5)--(w1)--(w2)--(w3)--(w4)--(w5);
	\end{scope}		
		
	\draw[->,very thick] (-2.5,0) -- (-1.5,0);		
		
	\node[H] (r) at (0:1.3) {};
	\node[H] (g) at (120:1.3) {};
	\node[H] (b) at (240:1.3) {};	
	\draw[He] (r.center)--(g.center) (g.center)--(b.center) (b.center)--(r.center);
	\node[H,dred] (r) at (0:1.3) {};	
	\node[H,dgreen] (g) at (120:1.3) {};
	\node[H,dblue] (b) at (240:1.3) {};			
\end{tikzpicture}

\begin{tikzpicture}[scale=0.8]

\begin{scope}[shift={(0,0)}]	
	\node[H] (r) at (0:1.3) {};
	\node[H] (g) at (120:1.3) {};
	\node[H] (b) at (240:1.3) {};	
	\draw[He] (r.center)--(g.center) (g.center)--(b.center) (b.center)--(r.center);
	\node[H,dred] (r) at (0:1.3) {};	
	\node[H,dgreen] (g) at (120:1.3) {};
	\node[H,dblue] (b) at (240:1.3) {};		
	\node[G] (v1) at ($(r)+(0:5pt)$) {};
	\node[G] (v2) at ($(g)+(100:5pt)$) {};
	\node[G1] (v3) at ($(b)+(200:4pt)$) {};
	\node[G] (v4) at ($(r)+(0:-6pt)$) {};
	\node[G] (v5) at ($(g)+(120:-6pt)$) {};			
	\draw[Ge] (v1)--(v2)--(v3)--(v4)--(v5)--(v1);
	
	\begin{scope}[shift={(0,3)}]
		\node[v,dgreen] (w5) at (0:1) {};	
		\node[v,dred] (w1) at (72:1) {};
		\node[v,dgreen] (w2) at (144:1) {};
		\node[v1,dblue] (w3) at (-144:1) {};
		\node[v,dred] (w4) at (-72:1) {};						
		\draw (w5)--(w1)--(w2)--(w3)--(w4)--(w5);
	\end{scope}
\end{scope}

\begin{scope}[shift={(3.3,0)}]	
	\node[H] (r) at (0:1.3) {};
	\node[H] (g) at (120:1.3) {};
	\node[H] (b) at (240:1.3) {};	
	\draw[He] (r.center)--(g.center) (g.center)--(b.center) (b.center)--(r.center);
	\node[H,dred] (r) at (0:1.3) {};	
	\node[H,dgreen] (g) at (120:1.3) {};
	\node[H,dblue] (b) at (240:1.3) {};		
	\node[G] (v1) at ($(r)+(0:5pt)$) {};
	\node[G] (v2) at ($(g)+(100:5pt)$) {};
	\node[G1] (v3) at ($(b)+(200:4pt)$) {};
	\node[G] (v4) at ($(r)+(0:-6pt)$) {};
	\node[G] (v5) at ($(b)+(120:-6pt)$) {};			
	\draw[Ge] (v1)--(v2)--(v3)--(v4)--(v5)--(v1);
	
	\begin{scope}[shift={(0,3)}]
		\node[v,dblue] (w5) at (0:1) {};	
		\node[v,dred] (w1) at (72:1) {};
		\node[v,dgreen] (w2) at (144:1) {};
		\node[v1,dblue] (w3) at (-144:1) {};
		\node[v,dred] (w4) at (-72:1) {};						
		\draw (w5)--(w1)--(w2)--(w3)--(w4)--(w5);
	\end{scope}	
\end{scope}

\begin{scope}[shift={(6.6,0)}]	
	\node[H] (r) at (0:1.3) {};
	\node[H] (g) at (120:1.3) {};
	\node[H] (b) at (240:1.3) {};	
	\draw[He] (r.center)--(g.center) (g.center)--(b.center) (b.center)--(r.center);
	\node[H,dred] (r) at (0:1.3) {};	
	\node[H,dgreen] (g) at (120:1.3) {};
	\node[H,dblue] (b) at (240:1.3) {};			
	\node[G] (v1) at ($(r)+(0:5pt)$) {};
	\node[G] (v2) at ($(g)+(100:5pt)$) {};
	\node[G1] (v3) at ($(b)+(200:4pt)$) {};
	\node[G] (v4) at ($(g)+(120:-7pt)$) {};
	\node[G] (v5) at ($(b)+(120:-6pt)$) {};			
	\draw[Ge] (v1)--(v2)--(v3)--(v4)--(v5)--(v1);
	
	\begin{scope}[shift={(0,3)}]
		\node[v,dblue] (w5) at (0:1) {};	
		\node[v,dred] (w1) at (72:1) {};
		\node[v,dgreen] (w2) at (144:1) {};
		\node[v1,dblue] (w3) at (-144:1) {};
		\node[v,dgreen] (w4) at (-72:1) {};						
		\draw (w5)--(w1)--(w2)--(w3)--(w4)--(w5);
	\end{scope}		
\end{scope}

\begin{scope}[shift={(9.9,0)}]	
	\node[H] (r) at (0:1.3) {};
	\node[H] (g) at (120:1.3) {};
	\node[H] (b) at (240:1.3) {};	
	\draw[He] (r.center)--(g.center) (g.center)--(b.center) (b.center)--(r.center);
	\node[H,dred] (r) at (0:1.3) {};	
	\node[H,dgreen] (g) at (120:1.3) {};
	\node[H,dblue] (b) at (240:1.3) {};		
	\node[G] (v1) at ($(r)+(0:5pt)$) {};
	\node[G] (v2) at ($(g)+(100:5pt)$) {};
	\node[G1] (v3) at ($(r)+(0:-6pt)$) {};
	\node[G] (v4) at ($(g)+(120:-7pt)$) {};
	\node[G] (v5) at ($(b)+(120:-6pt)$) {};			
	\draw[Ge] (v1)--(v2)--(v3)--(v4)--(v5)--(v1);
	
	\begin{scope}[shift={(0,3)}]
		\node[v,dblue] (w5) at (0:1) {};	
		\node[v,dred] (w1) at (72:1) {};
		\node[v,dgreen] (w2) at (144:1) {};
		\node[v1,dred] (w3) at (-144:1) {};
		\node[v,dgreen] (w4) at (-72:1) {};						
		\draw (w5)--(w1)--(w2)--(w3)--(w4)--(w5);
	\end{scope}		
\end{scope}
\end{tikzpicture}
	\vspace*{-1.5em}
	\caption{A sequence of 3-colorings of $C_5$ and the same sequence seen as a $K_3$-recoloring sequence of homomorphisms from $C_5$ to $K_3$ (a graph with three vertices: striped red, checkered green, dotted blue). One vertex of $C_5$ is thickened for clarity.}	
\label{fig:hom}			
\end{figure}
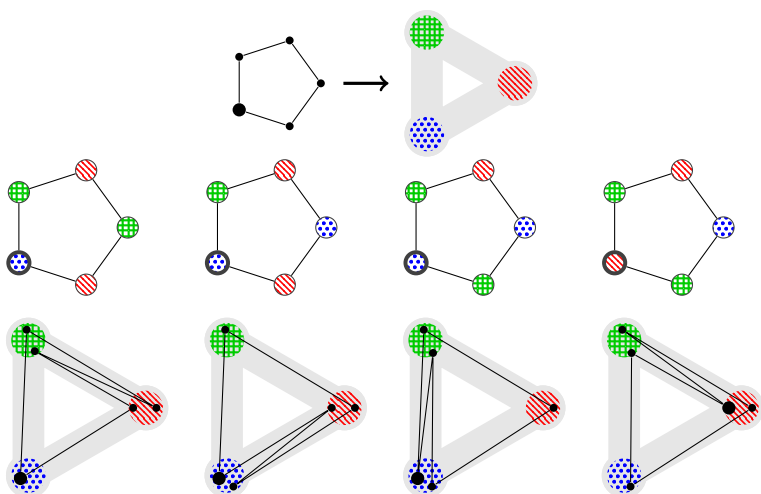

\paragraph*{Motivations}
The primary motivation for studying reconfiguration problems is to investigate the solution space of combinatorial problems, especially from the perspective of local search heuristics and random solution sampling. In particular, the success of a method for solving random CSPs called Survey Propagation is connected to several conjectures about the structure of clusters of satisfying assignments (connected components in the solution graph) and frozen variables (variables/vertices that cannot change their value/color by any sequence of reconfiguration steps), see~\cite{AchlioptasCR11}.

While finding paths in the solution graph is, for the above purposes, mostly a toy problem, it arises more directly in some settings, such as puzzles with sliding blocks. Indeed, the Nondeterministic Constraint Logic construction of Hearn and Demaine~\cite{HearnD05}, which gives a simple PSPACE-complete reconfiguration problem, allowed to show that many popular puzzles are \cclass{PSPACE}-complete~\cite{HearnD09,Mehta14a}. More interestingly, Heijltjes and Houston used the construction to prove that deciding the equivalence of proofs in a certain proof system is PSPACE-complete~\cite{Heijltjes2014}, answering a question about normal forms of proofs that arose in this context.

\paragraph*{Related work}
\problem{$H$-Recoloring} was shown to be \cclass{PSPACE}-complete (as a decision problem) for $H=K_k$ where $k\geq 4$ by Bonsma and Cereceda~\cite{BonsmaC09} and, surprisingly, in \cclass{P} for $k\leq 3$ by Cereceda et al.~\cite{CerecedaHJ11}. The latter result was improved by Jonhson et al.~\cite{johnson2014colouring} to show that \problem{Shortest $K_3$-Recoloring} is also in \cclass{P}.
Brewster et al.~\cite{BrewsterMMN16} recently generalized the dichotomy to circular colorings: when $H$ is a so called circular clique $K_{p/q}$ with $p/q<4$, then \problem{$H$-Recoloring} is in \cclass{P}, otherwise it is \cclass{PSPACE}-complete.

For CSPs in the Boolean domain, a dichotomy was shown by Gopalan et al.~\cite{GopalanKMP09}: for a fixed set of Boolean constraints $\Gamma$ (that is, Boolean relations, or \emph{clause types}), the problem of finding paths in the solution graph of a SAT$(\Gamma)$ instance is either in \cclass{P} or \cclass{PSPACE}-complete. In particular it is always in \cclass{P} when the corresponding satisfiability problem is in \cclass{P} (e.g. 2-SAT or Horn-SAT), but it is also in \cclass{P} for some $\Gamma$ for which satisfiability is \cclass{NP}-complete. This was slightly corrected (with a further correction in 2015) and extended to several similar problems by Schwerdtfeger~\cite{Schwerdtfeger14,Schwerdtfeger16}, while a trichotomy was shown for the problem of finding \emph{shortest} paths by Mouawad et al.~\cite{MouawadNPR15}.
Both \cite{GopalanKMP09} and \cite{Schwerdtfeger14} asked whether their results could be extended to larger domains.
Our work can be seen as a step in this direction, but limited to only one symmetric relation of arity 2.

The corresponding dichotomy for satisfiability, that is, deciding the existence of a solution, was proved by Schaefer~\cite{Schaefer78}. Generalizing it to CSPs with arbitrary finite domains is a long-standing open problem stated by Feder and Vardi~\cite{FederV98}. They showed that the conjecture is unchanged when limited to one relation of arity 2 (digraph homomorphisms).  Hell and Ne\v{s}et\v{r}il proved the dichotomy in the case the relation is additionally assumed to be symmetic (graph homomorphism)~\cite{HellN90}: the problem of deciding the existence of an $H$-coloring of a given graph is in \cclass{P} for $H$ bipartite or containing a loop, and \cclass{NP}-complete otherwise.

\paragraph*{Results}
It is natural to ask whether the unexpected tractability of \problem{$K_3$-Recoloring} (in light of \problem{3-Colorability} being \cclass{NP}-complete, even in graphs as simple as 4-regular planar graphs~\cite{dailey1980uniqueness}) is caused by the following property: whenever a vertex changes its color in a 3-coloring (e.g. from red to green), all of its neighbors must have one common color (not red nor green, hence blue).
We answer this in the positive by considering the following way to formalize this property: a graph $H$ (without loops) is \emph{square-free} if it does not contain a cycle on four vertices $C_4$ as a subgraph (not necessarily induced).
This is equivalent to requiring that for every two colors $a,b\in V(H)$, the set of common neighbors $N_H(a)\cap N_H(b)$ contains at most one color.\footnote{For graphs with loops allowed, this requirement is equivalent to excluding $C_4$, $K_3$ with one loop added, and $K_2$ with both loops added, as subgraphs. All results of this paper extend in a straightforward manner to graphs $H$ with loops allowed, using this as a definition of \emph{square-free}; a star graph with looped leaves gives an interesing example. We choose to omit the few additional group-theoretic details that would be needed in proofs.}
Note that $K_3$ as well as all graphs of girth at least 5 are square-free.

The main result is an algorithm that solves \problem{Shortest $H$-Recoloring} in polynomial time for all square-free graphs $H$, even if $H$ is given on input.
To achieve this, we characterize possible paths in $\Hom_1(G,H)$ (Theorem~\ref{thm:characterization}) by describing (in an exhaustive, but concise, algorithmic way: Theorem~\ref{thm:realizableEnumeration}) all sequences of colors which can occur when we observe one given vertex of $G$ throughout an $H$-recoloring.
It turns out they are mostly limited by the fact that reconfiguring an $H$-coloring corresponds to a continuous transformation of a map from $G$ to $H$ as topological spaces, an unexpected connection that might be interesting on its own.

\paragraph*{In combinatorial algebraic topology}
Reconfiguration of homomorphisms has already been studied independently in the field of combinatorial algebraic topology, though from a different angle.
$H$-recoloring, that is, reachability in the solution graph $\Hom_1(G,H)$ is equivalent to reachability in the so called \emph{exponential graph} $H^G$, or more exactly its subgraph induced by looped vertices.
This reachability relation of homomorphisms was also studied under the name \emph{$\times$-homotopy} by Dochtermann~\cite{Dochtermann09}.

The \emph{Hom-complex} $\Hom(G,H)$ is a construction slightly richer than $\Hom_1(G,H)$: it is the simplicial complex whose vertices are  $H$-colorings of $G$ and faces are those sets of $H$-colorings that can be arbitrarily mixed with each other (that is, any function which, on each vertex of $G$, agrees with some $H$-coloring from the set, has to be a valid $H$-coloring as well), see \cite{babson2006}.
Reachability in  $\Hom(G,H)$ is again equivalent to reachability in $\Hom_1(G,H)$.

The exponential graph $H^G$ was used to study properties of homomorphisms, especially in a category-theoretic setting, many applications are shown in~\cite{hell2004homomorphism_book}.
The Hom-complex was first used to provide lower bounds on the chromatic number of graphs, a notoriously hard problem, using topology.
A typical theorem derived from such methods is that for loopless graphs $G,H$, if $\Hom(G,H)$ (or equivalently, $\Hom_1(G,H)$) is connected for all $G$ of degree at most $d$, then the chromatic number of $H$ is at least $d/2$ (and is conjectured to be at least $d$) \cite{brightwell2004graph}.
Studies have thus been mostly concerned with highly regular graphs for which the Hom-complex can be proved to be in some sense tightly connected.

The computational complexity of deciding reachability of homomorphisms ($\times$-homotopy) has not been studied earlier, except for the clique and circular clique cases.
The characterization in this paper (Theorem~\ref{thm:characterization}) shows that it is tightly connected to the usual notion of homotopy (continuous transformations) of continuous maps corresponding to homomorphisms.
One direction (more precisely, the part formulated in Corollary~\ref{cor:topoValid}) should be unsurprising to readers familiar with combinatorial algebraic topology: one can assign topological spaces to graphs in a natural way, so that homomorphisms give rise to continuous maps and reconfiguration gives rise to homotopy (see e.g.~\cite{Dochtermann09}, Theorem 5.1.(2), for a formal, general statement).
The other direction, which allows us to infer (and even construct) a reconfiguration sequence essentially from a single homotopy of graph maps, is a new contribution.

\section{Preliminaries}
An (undirected) graph $G$ is a pair $(V(G),E(G))$ where $V(G)$ is a finite set of \emph{vertices},
while $E(G)$ is the set of \emph{edges}: undirected vertex pairs $\{u,v\}$, $u,v\in V(G)$, written $uv$ for short.
The neighborhood $N_G(v)$ is defined as $\{w \in V(G) \mid vw \in E(G)\}$.
A \emph{homomorphism} $\alpha: G\to H$, or $H$-coloring of $G$, is a function $V(G) \to V(H)$ such that $uv \in E(G)$ implies $\alpha(uv) := \alpha(u)\alpha(v) \in E(H)$.
\medskip

$G$ and $H$ in this paper are always connected undirected graphs with at least one edge and no loops.
\textbf{$H$ is always assumed to be square-free.}
\medskip

\pagebreak[3]

An \emph{$H$-recoloring sequence} or \emph{reconfiguration sequence} is a path in $\Hom_1(G,H)$, that is, a sequence of $H$-colorings of $G$ in which consecutive colorings differ at one vertex.
\problem{$H$-Recoloring} is the problem where given a graph $G$ and two $H$-colorings $\alpha,\beta$ of $G$, we ask whether there is an $H$-recoloring sequence between them (whether they are in the same connected component of $\Hom_1(G,H)$). In \problem{Shortest $H$-Recoloring} we are additionally given an integer $\ell$ and we ask whether there is an $H$-recoloring sequence of length at most $\ell$.
All our algorithms will work uniformly for $H$, that is, we can assume that a square-free graph $H$ is also part of the input.

Consider a step of an $H$-recoloring sequence: a vertex $v\in V(G)$ changes color from $a\in V(H)$ to $b\in V(H)$.
Since $G$ is connected, loopless and has an edge, $v$ has a neighbor, say $w\neq v$.
As only $v$ changes its color in the step, $w$ has the same color, say $h\in V(H)$, before and after the step.
The $H$-coloring before the step implies that $ha \in E(H)$, while the one after the step implies that $hb \in E(H)$.
Thus $h\in N_H(a) \cap N_H(b)$. 
From the assumption that $H$ is square-free we infer that $N_H(a) \cap N_H(b) = \{h\}$.
We will often call $h$ `\emph{the color that all neighbors of $v$ have during the step}' (that is, in the $H$-colorings just before and after the step), without arguing its existence and uniqueness anymore.

\subsection*{Fundamental groupoid}
We now define the fundamental groupoid of graph in elementary terms. The fundamental groupoid of a topological space is a basic tool of algebraic topology; discrete variants like the one defined here was also considered classically, the first chapter of~\cite{kwak2007graphs} provides an in-depth reference (including details on the case of graphs with loops).

An \emph{oriented edge} of a graph $H$ is an oriented pair $e=(u,v)$
such that $\{u,v\}$ is an edge of $H$;
we denote its initial vertex $u$ as $\iota(e)$ and its target vertex $v$ as $\tau(e)$.
We write $e^{-1}$ for $(\tau(e),\iota(e))$.
A \emph{walk} from $u$ to $v$ in a graph $H$ is a sequence of oriented edges $e_1 e_2 \dots e_l$ of $H$
such that $\iota(e_1)=u$, $\tau(e_l)=v$ and $\tau(e_i)=\iota(e_{i+1})$ for $i=0,\dots,l-1$.
We write $\varepsilon$ for an empty walk (though some define a different empty walk for every vertex, the endpoints of $\varepsilon$ will be irrelevant for us).
The length of a walk is the number of edges in it.
A walk $W_1$ from $u$ to $v$ can be concatenated to a walk $W_2$ from $v$ to $w$ to form a walk $W_1 W_2$ from $u$ to $w$.

\begin{figure}[H]
	\centering
	\begin{tikzpicture}[scale=0.7]
	\dfivegraph;
	
	\node[G1] (v1) at (ha2) {};
	\node[G] (v2) at (ha1) {};
	\node[G] (v3) at (ha0) {};
	\node[G1] (v4) at (ha4) {};
	\foreach \x [remember=\x as \lastx (initially 1)] in {2,...,4}
	{ \draw[Ge,->] (v\lastx)--(v\x); }	
	
\node at (2.1,0) {\Huge$\cdot$};	

\begin{scope}[shift={(4,0)}]
	\dfivegraph;
	
	\node[G1] (v1) at ($(ha4)-(-72:4pt)$) {};
	\node[G] (v2) at (ha0) {};
	\node[G] (v3) at (ha1) {};
	\node[G] (v4) at (ha2) {};
	\node[G] (v5) at (ha3) {};		
	\node[G1] (v6) at ($(ha4)+(-72:4pt)$) {};
	\foreach \x [remember=\x as \lastx (initially 1)] in {2,...,6}
	{ \draw[Ge,->] (v\lastx)--(v\x); }
\end{scope}

\node at (6.1,0) {\huge$=$};

\begin{scope}[shift={(8,0)}]
	\dfivegraph;
	
	\node[G1] (v1) at (ha2) {};
	\node[G] (v2) at (ha3) {};
	\node[G1] (v3) at (ha4) {};
	\foreach \x [remember=\x as \lastx (initially 1)] in {2,...,3}
	{ \draw[Ge,->] (v\lastx)--(v\x); }
\end{scope}

\begin{scope}[shift={(0,-3.5)}]

	\dfivegraph;
	
	\node[G1] (v1) at (ha2) {};
	\node[G] (v2) at (ha3) {};
	\node[G1] (v3) at (ha4) {};
	\foreach \x [remember=\x as \lastx (initially 1)] in {2,...,3}
	{ \draw[Ge,->] (v\lastx)--(v\x); }

\node at (2.1,0) {\Huge$\cdot$};

\begin{scope}[shift={(4,0)}]
	\dfivegraph;
	
	\node[G1] (v1) at ($(ha4)-(-72:4pt)$) {};
	\node[G] (v2) at (ha0) {};
	\node[G] (v3) at (ha1) {};
	\node[G] (v4) at (ha2) {};
	\node[G] (v5) at (ha3) {};		
	\node[G1] (v6) at ($(ha4)+(-72:4pt)$) {};
	\foreach \x [remember=\x as \lastx (initially 1)] in {2,...,6}
	{ \draw[Ge,->] (v\lastx)--(v\x); }
\end{scope}

\node at (6.1,0) {\huge$=$};

\begin{scope}[shift={(8,0)}]
	\dfivegraph;
	
	\node[G1] (v1) at ($(ha2)-(144:4pt)$) {};
	\node[G] (v2) at ($(ha3)-(-144:4pt)$) {};
	\node[G] (v3) at ($(ha4)-(-72:4pt)$) {};
	\node[G] (v4) at (ha0) {};
	\node[G] (v5) at (ha1) {};
	\node[G] (v6) at ($(ha2)+(144:4pt)$) {};
	\node[G] (v7) at ($(ha3)+(-144:4pt)$) {};
	\node[G1] (v8) at ($(ha4)+(-72:4pt)$) {};
	\foreach \x [remember=\x as \lastx (initially 1)] in {2,...,8}
	{ \draw[Ge,->] (v\lastx)--(v\x); }
\end{scope}	

\end{scope}	
\end{tikzpicture}
	\vspace*{-1.5em}
	\caption{Examples of $\cdot$ multiplication in the fundamental groupoid of $H=C_5$.} 
\label{fig:groupoid}
\end{figure}
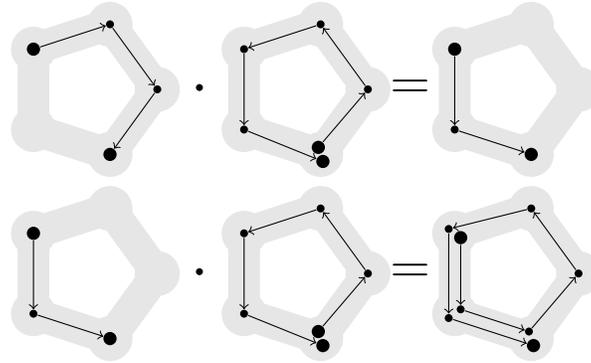

We call a walk \emph{reduced} if it contains no two consecutive edges $e_i e_{i+1}$ such that $e_{i+1}=e_i^{-1}$.
One can \emph{reduce} a walk by removing any such two consecutive edges from the sequence.
It can easily be seen that by iteratively reducing a walk $W$, one always gets the same reduced walk, which we denote as $\red{W}$, see Figure~\ref{fig:reducing}.
For any two reduced walks $\red{W_1},\red{W_2}$ such that $W_2$ starts where $W_1$ ends, we write $\red{W_1} \cdot \red{W_2}$ for $\red{\red{W_1}\ \red{W_2}}$ and similarly one can observe that $\cdot$ is associative.
For any walk $W=e_1 e_2\dots e_l$ we write $W^{-1}$ for the reversed walk $e_l^{-1}\dots e_2^{-1} e_1^{-1}$.
Clearly $\red{W} \cdot \red{W}^{-1} = \varepsilon = \red{W}^{-1} \cdot \red{W}$ and $\varepsilon \cdot \red{W} = \red{W} \cdot \varepsilon = \red{W}$, so the set of reduced walks of a graph forms together with the operations $\cdot$ and ${()}^{-1}$ a \emph{groupoid}, that is, it satifies all axioms of a group, except that the group operation $\cdot$ is a partial function, defined only when the `head' of one element matches the `tail' of the other. (A groupoid can also be defined as a category in which every morphism is invertible.) This particular groupoid is called the \emph{fundamental groupoid} $\pi(H)$ of $H$; see Figure~\ref{fig:groupoid}.

Groupoids behave similarly to groups (much more so than semigroups, for example) and identities such as $(e\cdot f)^{-1} = f^{-1} \cdot e^{-1}$ known from group theory are easily reproved in groupoids.
While we could define a group (the fundamental group) by considering only closed walks starting and ending in a chosen vertex $v$, this would make formulas less uniform, requiring some tedious additional steps when changing the base point $v$, for example.


\subsection*{Topological interpretation}
Let us comment on how this algebraic structure captures the topology of curves in the graph.
When referring to topology, continuous maps, curves and homotopy, we only give informal interpretations without proof. We recall some classical results and definitions, which we do not require formally, but are helpful, if not crucial, in understanding the results.

A graph $H$ without loops can be naturally associated with a topological space, constructed from copies of the unit interval $[0,1]\subseteq \mathbb{R}$ for each edge, with endpoints merged into vertices accordingly.
A \emph{curve} in this space is a continuous map $f:[0,1]\to H$ (where $H$ is meant as a topological space), not necessarily injective.
Two curves $f_0,f_1$ are \emph{homotopic} if one can be continuously transformed into the other, which means there is a set of functions $\{\phi_t : t\in [0,1]\}$ such that $\phi_0=f_0, \phi_1=f_1$ and the mapping $\phi: (t,x) \mapsto \phi_t(x)$ is continuous as a function from $[0,1]\times [0,1]$ to $H$.

The fundamental groupoid fully describes curves up to homotopy.
For any two vertices $u,v$ of $H$, two curves $f_0,f_1$ are homotopic via a homotopy $\phi_t$ that fixes the endpoints ($\phi_t(0)=u, \phi_t(1)=v$ for all $t$) if and only if the corresponding reduced walks in $\pi(H)$ are identical.
When no vertex is fixed, a closed curve starting and ending in $v_1$ is homotopic to a closed curve starting and ending in $v_2$ (via a homotopy such that $\phi_t(1)=\phi_t(0)$ for all $t$) if and only if the corresponding elements $C_1,C_2$ of $\pi(H)$ are \emph{conjugate}, meaning $C_2=P^{-1} \cdot C_1 \cdot P$ for some $P\in \pi(H)$.


\begin{figure}[H]
	\centering
	\begin{tikzpicture}[scale=0.8]
	\dtengraph;
	
	\node[G1] (v1) at (ha2) {};
	\node[G] (v2) at (ha1) {};
	\node[G] (v3) at ($(ha0)+(-3pt,7pt)$) {};
	\node[G] (v4) at ($(hb0)+(0pt,7pt)$) {};
	\node[G] (v5) at ($(hb4)+(-72:-4pt)$) {};
	\node[G] (v6) at ($(hb3)+(-144:-4pt)$) {};
	\node[G] (v7) at ($(hb2)+(144:-4pt)$) {};
	\node[G] (v8) at ($(hb1)+(72:-4pt)$) {};
	\node[G] (v9) at ($(hb0)+(2pt,-7pt)$) {};
	\node[G] (v10) at ($(ha0)+(2pt,-5pt)$) {};
	\node[G] (v11) at ($(hb0)+(4pt,-3pt)$) {};
	\node[G] (v12) at ($(hb1)+(72:4pt)$) {};
	\node[G] (v13) at ($(hb2)+(144:4pt)$) {};
	\node[G] (v14) at ($(hb3)+(-144:4pt)$) {};
	\node[G] (v15) at ($(hb4)+(-72:4pt)$) {};
	\node[G] (v16) at ($(hb0)+(2pt,2pt)$) {};
	\node[G] (v17) at ($(ha0)+(-3pt,-0pt)$) {};
	\node[G1] (v18) at (ha4) {};					
	\foreach \x [remember=\x as \lastx (initially 1)] in {2,...,18}
	{ \draw[Ge,->] (v\lastx)--(v\x); }
	
\begin{scope}[shift={(7.5,0)}]
	\dtengraph;

	\node[G1] (v1) at (ha2) {};
	\node[G] (v2) at ($(ha1)+(0pt,-5pt)$) {};
	\node[G] (v3) at ($(ha0)+(-6pt,1pt)$) {};
	\node[G] (v4) at ($(ha1)+(2pt,4pt)$) {};
	\node[G] (v5) at ($(ha0)+(72:4pt)$) {};
	\node[G] (v6) at ($(hb0)+(0,4pt)$) {};
	\node[G] (v7) at ($(hb4)+(108:5pt)$) {};
	\node[G] (v8) at ($(hb3)+(72:4pt)$) {};
	\node[G] (v9) at ($(hb4)+(108:0pt)$) {};
	\node[G] (v10) at ($(hb0)+(4pt,-1pt)$) {};
	\node[G] (v11) at ($(hb1)+(-108:-4pt)$) {};
	\node[G] (v12) at ($(hb2)+(-36:-4pt)$) {};
	\node[G] (v13) at ($(hb3)+(-3pt,-4pt)$) {};
	\node[G] (v14) at ($(hb4)+(128:-6pt)$) {};
	\node[G] (v15) at ($(hb3)+(4pt,-2pt)$) {};
	\node[G] (v16) at ($(hb2)+(-36:4pt)$) {};
	\node[G] (v17) at ($(hb1)+(-108:4pt)$) {};
	\node[G] (v18) at ($(hb0)+(-2pt,-5pt)$) {};	
	\node[G] (v19) at ($(ha0)+(2pt,-5pt)$) {};	
	\node[G] (v20) at ($(ha4)+(-72:4pt)$) {};	
	\node[G] (v21) at ($(ha3)+(0pt,-0pt)$) {};		
	\node[G1] (v22) at ($(ha4)+(-72:-4pt)$) {};					
	\foreach \x [remember=\x as \lastx (initially 1)] in {2,...,22}
	{ \draw[Ge,->] (v\lastx)--(v\x); }	
\end{scope}

\begin{scope}[shift={(0,-3.6)}]
	\dtengraph;
	
	\node[G1] (v1) at (ha2) {};
	\node[G] (v2) at (ha1) {};
	\node[G] (v3) at (ha0) {};
	\node[G1] (v4) at (ha4) {};
	\foreach \x [remember=\x as \lastx (initially 1)] in {2,...,4}
	{ \draw[Ge,->] (v\lastx)--(v\x); }
		
\begin{scope}[shift={(7.5,0)}]
	\dtengraph;
	
	\node[G1] (v1) at (ha2) {};
	\node[G] (v2) at (ha1) {};
	\node[G] (v3) at ($(ha0)+(0pt,4pt)$) {};
	\node[G] (v4) at ($(hb0)+(0pt,4pt)$) {};
	\node[G] (v5) at (hb4) {};
	\node[G] (v6) at (hb3) {};
	\node[G] (v7) at (hb2) {};
	\node[G] (v8) at (hb1) {};
	\node[G] (v9) at ($(hb0)-(0pt,4pt)$)  {};						
	\node[G] (v10) at ($(ha0)-(0pt,4pt)$)  {};							
	\node[G1] (v11) at (ha4) {};
	\foreach \x [remember=\x as \lastx (initially 1)] in {2,...,11}
	{ \draw[Ge,->] (v\lastx)--(v\x); }
\end{scope}

\end{scope}

\draw[thick,->,decorate,decoration=snake] (0.6,-1) -- (0.6,-2.1);
\draw[thick,->,decorate,decoration=snake] (4.7,-1) -- (3.90,-2.3);

%
%
%
\end{tikzpicture}
	\vspace*{-1.5em}	
	\caption{Examples of two walks (in a 'dumbbell' graph $H$ on 10 vertices) which reduce to the same, bottom left one. The bottom right one is a different reduced walk; when its endpoints are fixed, it cannot be distorted as a curve to give any of the others.}	
\label{fig:reducing}	
\end{figure}
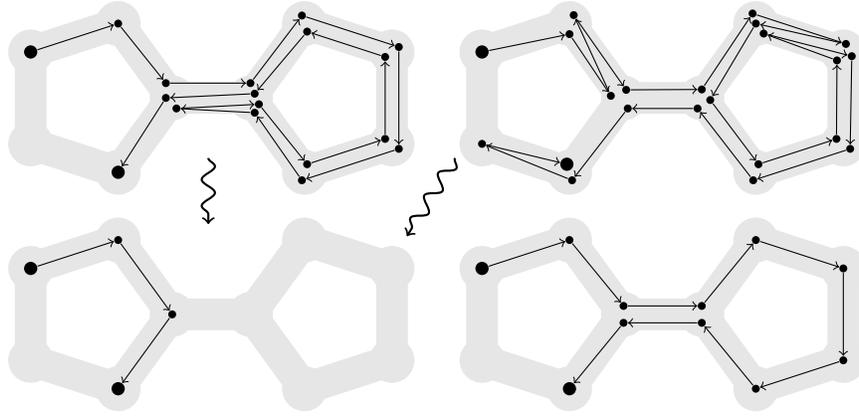

\section{Vertex walks and realizability}\label{sec:realizability}
As a vertex $v\in V(G)$ changes colors from $a$ to $b$ in a step of a $H$-recoloring sequence, let $h$ be the color that all neighbors of $v$ have during the change. Then $(a,h)(h,b)$ is a walk in $H$ (of length 2). Looking at all the color changes of one vertex this way gives a walk in $H$ which traces the colors that $v$ had. This walk (for an arbitrarily chosen vertex $v$), even after reducing, will be shown to almost completely describe the $H$-recoloring sequence.

Formally, consider an $H$-recoloring sequence $S=\sigma_0,\dots,\sigma_l$ of $G$ and any vertex $v\in V(G)$. We define $S(v)$ as the following walk in $H$.
If $l=0$ ($S$ is empty), then $S(v)=\varepsilon$.
If $l=1$ ($S$ contains only one reconfiguration step) then
$S(v) = \varepsilon$ when $\sigma_0(v)=\sigma_1(v)$ and
$S(v) = (\sigma_0(v),h) (h,\sigma_1(v))$ otherwise,
$h$ being the color that all neighbors of $v$ have in $\sigma_0$ and $\sigma_1$.
If $l>1$, then $S(v)=S_0(v) S_1(v) \dots S_{l-1}(v)$,
where $S_i$ is the subsequence $\sigma_i,\sigma_{i+1}$ of $S$.

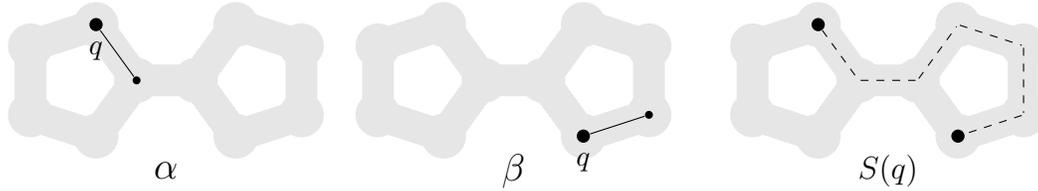
\begin{figure}[H]
	\centering
	\begin{tikzpicture}[scale=0.6]
\begin{scope}[yscale=-1]
	\dtengraph;	
	\node[G1,label=below:$q$] (v0) at (ha4) {};
	\node[G] (v1) at (ha0) {};	
	\draw[Ge] (v0)--(v1);
	\node at (0.65,+2) {\Large$\alpha$};
	
\begin{scope}[shift={(7.7,0)}]
	\dtengraph;
	\node[G1,label=below:$q$] (v0) at (hb4) {};
	\node[G] (v1) at (hb3) {};	
	\draw[Ge] (v0)--(v1);	
	\node at (0.65,+2) {\Large$\beta$};	
\end{scope}	

\begin{scope}[shift={(16,0)}]
	\dtengraph;
	\node[G1] (v0) at (ha4) {};
	\node[Gv] (v1) at (ha0) {};	
	\node[Gv] (v2) at (hb0) {};	
	\node[Gv] (v3) at (hb1) {};	
	\node[Gv] (v4) at (hb2) {};	
	\node[Gv] (v5) at (hb3) {};	
	\node[G1] (v6) at (hb4) {};
	\draw[Gve] (v0)--(v1)--(v2)--(v3)--(v4)--(v5)--(v6);
	\node at (0.65,+2) {\large$S(q)$};		
\end{scope}	
\end{scope}
\end{tikzpicture}
	\vspace*{-1.5em}
	\caption{A realizable walk for $\alpha,\beta:K_2\to H$ and $q$.
	Note that the shortest walk from $\alpha(q)$ to $\beta(q)$ (of length $3$) is not realizable because of parity.}
\label{fig:vertexWalk}
\end{figure}

For two $H$-coloring $\alpha,\beta$ of $G$ and an arbitrarily fixed vertex $q\in V(G)$, we call a reduced walk $Q\in\pi(H)$ from $\alpha(q)$ to $\beta(q)$ a \emph{realizable} walk if there is an $H$-recoloring sequence $S=\sigma_0,\dots,\sigma_l$ such that $\sigma_0=\alpha, \sigma_l=\beta$ and $\red{S(q)}=Q$. Instead of asking just whether any sequence exists, we focus on the following question: which elements of $\pi(H)$ are realizable? In other words: which walks in $H$ can be realized, up to reductions, as vertex walks of $q$ in some solution sequence. It is immediate from the definition that $Q$ must have even length (notice that the parity of the length of walks is preserved by reducing, since we only remove pairs of edges $ee^{-1}$), see Figure~\ref{fig:vertexWalk}.

\pagebreak

Parity is one of three conditions that characterize realizable walks. Sections~\ref{sec:topo}, \ref{sec:tight} describe the second (topological) and third necessary conditions, respectively. 
In Section~\ref{sec:char} we prove they are sufficient (the characterization), Section~\ref{sec:calculations} describes algorithmically the topological condition, and finally Section~\ref{sec:algo} uses these to give the main algorithm.
\medskip

Apart from vertex walks, another kind of walk in $H$ we will often see is the following.
If $W=e_1 \dots e_\ell$ is a walk in $G$ and $\alpha$ is an $H$-coloring of $G$, then observe that $\alpha(W) := \alpha(e_1) \dots \alpha(e_\ell)$ is a walk in $H$ (of the same length).

\section{Topological validity}\label{sec:topo}
With a homomorphism from $G$ to $H$ one can associate a continuous map from $G$ to $H$ (understood as topological spaces, as described in Preliminaries).
Intuitively, a reconfiguration between two $H$-colorings then corresponds to a homotopy (a continuous transformation) between the two corresponding maps.
While much less constrained, homotopies still must preserve certain invariants.

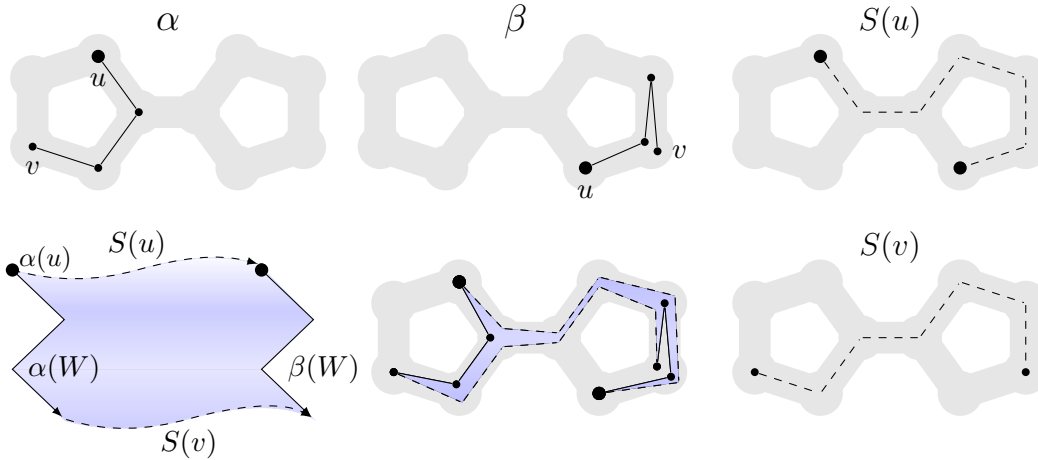
\begin{figure}[H]
	\centering
	\begin{tikzpicture}[scale=0.6]
\begin{scope}[yscale=-1]
	\dtengraph;	
	\node[G1,label=below:$u$] (v0) at (ha4) {};
	\node[G] (v1) at (ha0) {};	
	\node[G] (v2) at (ha1) {};		
	\node[G,label=below:$v$] (v3) at (ha2) {};
	\draw[Ge] (v0)--(v1)--(v2)--(v3);
	\node at (0.65,-2) {\Large$\alpha$};
	
\begin{scope}[shift={(7.7,0)}]
	\dtengraph;
	\node[G1,label=below:$u$] (v0) at (hb4) {};
	\node[G] (v1) at ($(hb3)-(36:5pt)$) {};	
	\node[G] (v2) at (hb2) {};
	\node[G,label=right:$v$] (v3) at ($(hb3)+(36:5pt)$) {};	
	\draw[Ge] (v0)--(v1)--(v2)--(v3);	
	\node at (0.65,-2) {\Large$\beta$};	
\end{scope}	

\begin{scope}[shift={(16,0)}]
	\dtengraph;
	\node[G1] (u0) at (ha4) {};
	\node[Gv] (u1) at (ha0) {};	
	\node[Gv] (u2) at (hb0) {};	
	\node[Gv] (u3) at (hb1) {};	
	\node[Gv] (u4) at (hb2) {};	
	\node[Gv] (u5) at (hb3) {};	
	\node[G1] (u6) at (hb4) {};
	\draw[Gve] (u0)--(u1)--(u2)--(u3)--(u4)--(u5)--(u6);
	\node at (0.65,-2) {\large$S(u)$};		
\end{scope}	

\begin{scope}[shift={(16,5)}]
	\dtengraph;
	\node[G] (v0) at (ha2) {};
	\node[Gv] (v1) at (ha1) {};	
	\node[Gv] (v2) at (ha0) {};	
	\node[Gv] (v3) at (hb0) {};	
	\node[Gv] (v4) at (hb1) {};	
	\node[Gv] (v5) at (hb2) {};	
	\node[G] (v6) at (hb3) {};
	\draw[Gve] (v0)--(v1)--(v2)--(v3)--(v4)--(v5)--(v6);
	\node at (0.65,-2) {\large$S(v)$};		
\end{scope}	

\begin{scope}[shift={(8,5)}]
	\dtengraph;
	\node[G1] (a0) at (ha4) {};
	\node[G] (a1) at ($(ha0)-(0:6pt)$) {};	
	\node[G] (a2) at ($(ha1)-(72:6pt)$) {};		
	\node[G] (a3) at (ha2) {};
	\draw[Ge] (a0)--(a1)--(a2)--(a3);	
	\node[G1] (b0) at (hb4) {};
	\node[G] (b1) at ($(hb3)+(36:5pt)$) {};	
	\node[G] (b2) at (hb2) {};
	\node[G] (b3) at ($(hb3)-(36:6pt)$) {};	
	\draw[Ge] (b0)--(b1)--(b2)--(b3);		
	\node[G1] (u0) at (ha4) {};
	\node[Gv] (u1) at ($(ha0)+(-72:6pt)$) {};	
	\node[Gv] (u2) at ($(hb0)+(0,-3pt)$) {};	
	\node[Gv] (u3) at ($(hb1)+(0,-3pt)$) {};	
	\node[Gv] (u4) at ($(hb2)+(-36:8pt)$) {};	
	\node[Gv] (u5) at ($(hb3)+(36:11pt)$) {};	
	\node[G1] (u6) at (hb4) {};
	\draw[Gve] (u0)--(u1)--(u2)--(u3)--(u4)--(u5)--(u6);
	\node[G] (v0) at (ha2) {};
	\node[Gv] (v1) at ($(ha1)+(72:6pt)$) {};	
	\node[Gv] (v2) at ($(ha0)+(72:6pt)$) {};	
	\node[Gv] (v3) at ($(hb0)+(0,3pt)$) {};	
	\node[Gv] (v4) at ($(hb1)+(0,3pt)$) {};	
	\node[Gv] (v5) at ($(hb2)-(-36:7pt)$) {};	
	\draw[Gve] (v0)--(v1)--(v2)--(v3)--(v4)--(v5)--(b3);
	\fill[draw=none,shade,top color=blue!25,bottom color=blue!25,middle color=blue!15] (ha4.center) to (a1.center) to (a2.center) to (ha2.center) to (v1.center) to (v2.center) to (v3.center) to (v4.center) to (v5.center) to (b3.center) to (b2.center) to (b1.center) to (hb4.center) to (u5.center) to (u4.center) to (u3.center) to (u2.center) to (u1.center) to (ha4.center);

	\node[G1] (a0) at (ha4) {};
	\node[G] (a1) at ($(ha0)-(0:6pt)$) {};	
	\node[G] (a2) at ($(ha1)-(72:6pt)$) {};		
	\node[G] (a3) at (ha2) {};
	\draw[Ge] (a0)--(a1)--(a2)--(a3);	
	\node[G1] (b0) at (hb4) {};
	\node[G] (b1) at ($(hb3)+(36:5pt)$) {};	
	\node[G] (b2) at (hb2) {};
	\node[G] (b3) at ($(hb3)-(36:6pt)$) {};	
	\draw[Ge] (b0)--(b1)--(b2)--(b3);		
	\node[G1] (u0) at (ha4) {};
	\node[Gv] (u1) at ($(ha0)+(-72:6pt)$) {};	
	\node[Gv] (u2) at ($(hb0)+(0,-3pt)$) {};	
	\node[Gv] (u3) at ($(hb1)+(0,-3pt)$) {};	
	\node[Gv] (u4) at ($(hb2)+(-36:8pt)$) {};	
	\node[Gv] (u5) at ($(hb3)+(36:11pt)$) {};	
	\node[G1] (u6) at (hb4) {};
	\draw[Gve] (u0)--(u1)--(u2)--(u3)--(u4)--(u5)--(u6);
	\node[G] (v0) at (ha2) {};
	\node[Gv] (v1) at ($(ha1)+(72:6pt)$) {};	
	\node[Gv] (v2) at ($(ha0)+(72:6pt)$) {};	
	\node[Gv] (v3) at ($(hb0)+(0,3pt)$) {};	
	\node[Gv] (v4) at ($(hb1)+(0,3pt)$) {};	
	\node[Gv] (v5) at ($(hb2)-(-36:7pt)$) {};	
	\draw[Gve] (v0)--(v1)--(v2)--(v3)--(v4)--(v5)--(b3);       
\end{scope}	
\end{scope}

\begin{scope}[shift={(-2.8,-3.5)},xscale=-2.3,yscale=1.1,rotate=-90]
 \node[inner sep=0pt] (b0) at (2,0) {};
 \node[inner sep=0pt] (b1) at (2,-2.4) {};
 \node[inner sep=0pt] (c0) at (3,-0.5) {};
 \node[inner sep=0pt] (c1) at (3,-2.9) {};
 \node[G1] (a1) at (0,-2.4) {};  
  
 \path[draw=blue!10,fill,shade,top color=white,bottom color=blue!10,middle color=blue!20] (0,0) to[out=-60,in=60] (0,-1.2) to[out=-120,in=120] (a1)
	 to (1,-2.9) to (b1.center)
	 to (2,-1.2) to (b0.center)
	 to (1,-0.5) to (0,0);  
 \path[draw=blue!10,fill,shade,top color=blue!10,bottom color=blue!20] (b0.center) to (2,-1.2) to (b1.center)
	 to (c1.center)
	 to[out=140,in=-120] (3,-1.7) to[out=60,in=-60] (c0)
	 to (b0.center);
	 
 \node[G1,label={[shift={(-4pt,4pt)}]right:\small$\alpha(u)$}] at (0,0) {};	 
 \node at (-.5,-1.2) {$S(u)$};
 \node at (3.5,-1.7) {$S(v)$}; 
 \node at (2,-0.5) {$\alpha(W)$};  
 \node at (2,-3) {$\beta(W)$};  	 
 \node[G1] (a1) at (0,-2.4) {}; 
	 
 \draw[Gve,->,arrows={-latex}] (0,0) to[out=-60,in=60] (0,-1.2) to[out=-120,in=120] (a1);
 \draw[Ge,->,arrows={-latex}] (0,-2.4) to (1,-2.9) to (2,-2.4) to (c1);
 \draw[Gve,->,arrows={-latex}] (c0) to[out=-60,in=60] (3,-1.7) to[out=-120,in=140] (c1); 
 \draw[Ge,->,arrows={-latex}] (0,0) to (1,-0.5) to (2,0) to (c0);
\end{scope}
\end{tikzpicture}
	\vspace*{-2em}
	\caption{
	Intuitively, if $\alpha$ can be transformed to $\beta$ by reconfiguration, then it can by a homotopy $\phi:[0,1]\times[0,|W|]\to H$ such that $\phi(0,\cdot) = \alpha(W)$ and $\phi(1,\cdot)=\beta(W)$.
	Let $S(u)=\phi(\cdot,0)$ and $S(v)=\phi(\cdot,|W|)$. Since $\phi$ is a continuous mapping of a rectangle to $H$ and since the boundary of the rectangle can be contracted to a point, the image of this boundary can also be contracted: 
	$\red{\alpha(W)}^{-1}\cdot \red{S(u)} \cdot \red{\beta(W)} \cdot \red{S(v)}^{-1} = \varepsilon$.}
\label{fig:topology}	
\end{figure}

We can describe essentially all of these invariants by considering the reconfiguration of an $H$-coloring of walk $W$ from $u$ to $v$ in $G$, which corresponds to continuously transforming a curve in $H$. 
The following lemma states in simple algebraic terms a key implication of this continuity:
the path traced by one endpoint $v$ of the curve is (up to reductions) the same as the following path: first going to the other endpoint $u$ along the initial curve's position $\alpha(W)$ in $H$, then tracing $u$, and then going back to $v$ along the curve's final position $\beta(W)$ in $H$, see Figure~\ref{fig:topology}.

\begin{lemma} 
\label{lem:onePathImpliesAll}
	Let $S=\sigma_0,\dots,\sigma_l$ be an $H$-recoloring sequence of $G$ from $\alpha=\sigma_0$ to $\beta=\sigma_l$.
	Consider any walk $W$ from vertex $u$ to $v$ in $G$.
	Then $\red{S(v)} = \red{\alpha(W)}^{-1} \cdot  \red{S(u)} \cdot \red{\beta(W)}$.
\end{lemma}

\begin{proof}
	The proof uses induction and the square-free property of $H$ for the base case.
	Assume first that $l=1$, so $S=\sigma_0,\sigma_1$, where $\sigma_1$ is obtained from $\sigma_0$ by recoloring one vertex $w\in V(G)$ from $\sigma_0(w)=a$ to $\sigma_1(w)=b$.
	Let $h$ be the color that all neighbors of $w$ have in $\sigma_0$ and $\sigma_1$. By definition of vertex walks, $S(w)=(a,h)(h,b)$ and all other vertex walks are empty.
	
	If $W=\varepsilon$, then $u=v$ and the claim follows trivially.
	
	If $W$ has length one, that is $W=(u,v)$, then one of the following holds:
	\begin{itemize}[itemsep=4pt,topsep=4pt,partopsep=4pt,parsep=0pt]
	\item $u\neq w$ and $v\neq w$, implying\\
	\hspace*{1.2em}	$S(u)=S(v)=\varepsilon$ and $\sigma_0(W)=\sigma_l(W)$.
	\item $u \neq w$ and $v=w$, implying\\
	\hspace*{1.2em} $S(u)=\varepsilon$, $S(v)=(a,h)(h,b)$ and $\sigma_0(W)=(h,a)$, $\sigma_l(W)=(h,b)$.
	\item $u=w$ and $v\neq w$, implying\\
	\hspace*{1.2em} $S(u)=(a,h)(h,b)$, $S(v)=\varepsilon$ and $\sigma_0(W)=(a,h)$, $\sigma_l(W)=(b,h)$.
	\end{itemize}	
	In each case 
		$\red{\sigma_l(W)} = \red{S(u)}^{-1} \cdot \red{\sigma_0(W)} \cdot \red{S(v)},$
	which is equivalent to the claim.
	
	If $W$ is longer, the claim follows inductively by splitting it into shorter walks: $W=W_1 W_2$ for 
	$W_1$ from $u$ to some vertex $w$ and $W_2$ from $w$ to $v$.
	It then follows that $\red{\sigma_l(W)} =$
	$$= \red{\sigma_l(W_1)} \cdot \red{\sigma_l(W_2)} =$$
	$$= \red{S(u)}^{-1} \cdot \red{\sigma_0(W_1)} \cdot \red{S(w)} \cdot
	    \red{S(w)}^{-1} \cdot \red{\sigma_0(W_2)} \cdot \red{S(v)} =$$
	$$= \red{S(u)}^{-1} \cdot \red{\sigma_0(W_1)} \cdot \red{\sigma_0(W_2)} \cdot \red{S(v)} =$$
	$$= \red{S(u)}^{-1} \cdot \red{\sigma_0(W)} \cdot \red{S(v)}.$$

	It remains to consider the case where $S$ has more than one step.
	Then the claim follows inductively by writing $S$ as the concatenation of shorter sequences:
	$S_1$ which ends in $\sigma_i$ and $S_2$ which starts in $\sigma_i$.
	It then follows that $S(v)=S_1(v)S_2(v)$ and $\red{S(v)} =$
	$$= \red{S_1(v)} \cdot \red{S_2(v)} = $$
	$$= \red{\sigma_0(W)}^{-1} \cdot  \red{S_1(u)} \cdot \red{\sigma_i(W)} \cdot 	\red{\sigma_i(W)}^{-1} \cdot  \red{S_2(u)} \cdot \red{\sigma_l(W)} =$$
	$$= \red{\sigma_0(W)}^{-1} \cdot  \red{S_1(u)} \cdot  \red{S_2(u)} \cdot \red{\sigma_l(W)} =$$
	\vspace{\abovedisplayshortskip}
	\hfill $\displaystyle= \red{\sigma_0(W)}^{-1} \cdot  \red{S(u)} \cdot \red{\sigma_l(W)}.$
\end{proof}

The lemma has two important corollaries.
\smallskip

First, in a given instance of \problem{$H$-Recoloring}, the reduced vertex walk $\red{S(q)}$ of one vertex $q$ in a $H$-recoloring sequence $S$ determines up to reductions (in other words, up to homotopy) all other vertex walks (since $\alpha$ and $\beta$ are given).
Later we will see that in shortest solution sequences all vertex walks are already reduced, so $\red{S(q)}$ actually determines the sequence exactly, up to reordering color changes of different vertices.
This means shortest recoloring sequences can be concisely represented by one realizable element $Q\in \pi(H)$ (the possible reorderings will be revealed in the proof of the characterization theorem).
This is also the reason for which we can focus on one walk and its realizability, instead of trying to describe an entire recoloring sequence.
\smallskip

Second, observe that the equality in the lemma holds for all walks in $G$, even though different walks between the same endpoints could \emph{a priori} give different values.
For every closed walk $C$ from $v$ to $v$ in $G$, we infer some equation on $\red{S(v)}$, namely $\red{S(v)} = \red{\alpha(C)}^{-1} \cdot  \red{S(v)} \cdot \red{\beta(C)}$, which expresses a certain topological condition on how solution sequences look like. 
We can rearrange this condition as 
$$ \red{S(v)}^{-1} \cdot \red{\alpha(C)} \cdot \red{S(v)} = \red{\beta(C)}  $$
In group theory we say that $\red{\alpha(C)}$ and $\red{\beta(C)}$ are \emph{conjugate} and that $\red{S(v)}$ is a witness of that.
We say a walk is \emph{topologically valid} if it satisfies the above equation for all $C$:

\begin{definition}
	\label{def:topoValidDef}
	Let $\alpha,\beta$ be two $H$-colorings of $G$ and let $q$ be a vertex of $G$.
	A walk $Q \in \pi(H)$ is \emph{topologically valid} for $\alpha,\beta,q$
	if for every closed walk $C$ from $q$ to $q$ we have
		$\red{\beta(C)} = Q^{-1} \cdot \red{\alpha(C)} \cdot Q$.		
\end{definition}

\begin{corollary}
\label{cor:topoValid}
	If~$Q\in \pi(H)$ is realizable for $\alpha,\beta,q$
	then $Q$ is topologically valid for $\alpha,\beta,q$.
\end{corollary}

We analyze such conjugacy equations in more detail in Section~\ref{sec:calculations}, for now let us give their intuitive meaning.
The condition that $\red{\alpha(C)}$  and  $\red{\beta(C)}$ are conjugate means that $C$, during any reconfiguration, always maps around the same cycle in $H$, up to reductions and rotations. (In the language of topology, $\red{\alpha(C)}$  and  $\red{\beta(C)}$ are homotopic, via homotopies that do not necessarily fix the base point $v$ of $C$).
The number of times the image of $C$ winds around this cycle in $H$ must also remain unchanged (in the special case $H=K_3$ this is exactly one of the conditions for 3-recoloring given by~\cite[Theorem 7 (C2)]{CerecedaHJ11}).
Finally, the condition that the realized walk $\red{S(v)}$ must be a witness will imply that two realizable walks (two solutions) can differ only in the number of times they wind around this cycle, essentially.
\medskip

In the next, final lemma of this section, we show that equations for closed walks already imply all other equations that would follow from Lemma~\ref{lem:onePathImpliesAll}.
In fact, one could show that $Q$ is topologically valid for $\alpha,\beta,q$ if and only if there is a homotopy continuously transforming $\alpha$ to $\beta$ such that $q$ traces the curve $Q$ throughout this transformation (that is, $\phi_0=\alpha, \phi_1=\beta$ and the image of $t\mapsto \phi_t(q)$ is $Q$).
This means that Corollary~\ref{cor:topoValid} is the strongest we can achieve using only this topological setting.

\begin{lemma}\label{lem:oneImpliesAll}
	If a walk $Q \in \pi(H)$ is \emph{topologically valid} for $\alpha,\beta,q$,
	then for any vertex $v$ and any two walks $W_1,W_2$ from $q$ to $v$ in $G$ we have 
	$\red{\alpha(W_1)}^{-1} \cdot  Q \cdot \red{\beta(W_1)} = \red{\alpha(W_2)}^{-1} \cdot  Q \cdot \red{\beta(W_2)}$.
\end{lemma}
\begin{proof}
	$W_1W_2^{-1}$ is a closed walk starting and ending in $q$, so 
	$\red{\beta(W_1W_2^{-1})} = Q^{-1} \cdot \red{\alpha(W_1W_2^{-1})} \cdot Q$.
	Therefore
		$$\red{\alpha(W_2)}\cdot \red{\alpha(W_1)}^{-1} \cdot Q \cdot \red{\beta(W_1)} \cdot \red{\beta(W_2)}^{-1} =$$
		$$= \red{\alpha(W_1W_2^{-1})}^{-1} \cdot Q \cdot \red{\beta(W_1W_2^{-1})} =$$
		$$= \red{\alpha(W_1W_2^{-1})}^{-1} \cdot Q \cdot Q^{-1} \cdot \red{\alpha(W_1W_2^{-1})} \cdot Q =$$	
		$$= Q.$$
Left-multiplying the equation by $\red{\alpha(W_2)}^{-1}$ and right-multiplying by $\red{\beta(W_2)}$ gives the claim.
\end{proof}



\section{Tight closed walks and frozen vertices}\label{sec:tight}
There is one more necessary condition for a walk to be realizable, beside even length and topological validity: intuitively, closed walks that map to walks tightly stretched around $H$ cannot be reconfigured in any way.

Formally, in an $H$-coloring $\alpha$ of $G$, a vertex $v$ of $G$ is called \emph{frozen} if for every $H$-recoloring sequence from $\alpha$ the resulting $H$-coloring $\beta$ has $\beta(v)=\alpha(v)$.
A closed walk $C=e_1 e_2\dots e_l$ is \emph{cyclically reduced} if it is reduced ($e_i \neq e_{i+1}^{-1}$) and additionally $e_l \neq e_1^{-1}$. In other words, repeating $C$ gives an infinite reduced walk. 
A closed walk $C$ is $\alpha$-tight if $\alpha(C)$ is cyclically reduced.

\begin{lemma}
\label{lem:tightAreFrozen}
	Let $\alpha$ be an $H$-coloring of $G$ and let $C$ be an $\alpha$-tight walk in $G$.
	Then all vertices of $C$ are frozen in $\alpha$.
\end{lemma}
\begin{proof}
	Suppose to the contrary that there is an $H$-recoloring sequence $\sigma_0,\dots,\sigma_l$ from $\alpha$, such that $\sigma_l(C)\neq \sigma_0(C)$.
	Let $i$ be the least such that $\sigma_i(C)\neq \sigma_0(C)$.
	Then in $\sigma_{i-1}$ all vertices of $C$ have the same color as in $\alpha=\sigma_0$, so $\sigma_{i-1}(C)$ is cyclically reduced, while $\sigma_i$ is obtained from $\sigma_{i-1}$ by changing the color of some vertex $v\in C$ from $a$ to $b$. Let $h$ be the color that all neighbors of $v$ have in $\sigma_{i-1}$ and $\sigma_i$.
	Let $u,w$ be the vertices of $C$ just before and just after $v$ on $C$.
	Since they are neighbors of $v$, they must have the color $h$ in $\sigma_{i-1}$.
	But then $\sigma_{i-1}$ maps the subsequent edges $(u,v)(v,w)$ of $C$ to $(h,a)(a,h)$, contradicting that $\sigma_{i-1}(C)$ is cyclically reduced.
\end{proof}

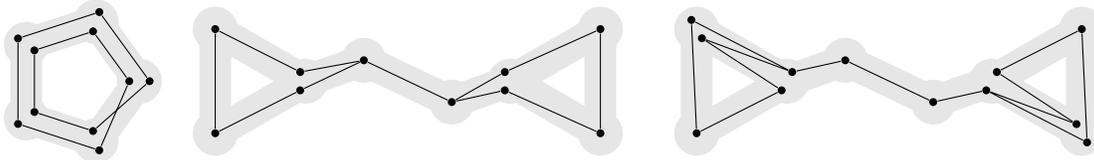
\begin{figure}[H]
	\centering
	\begin{tikzpicture}[scale=0.8]
\begin{scope}[shift={(-5.5,0)},scale=0.8]
	\dfivegraph;
	\node[G] (v0) at ($(ha0)+(0:6pt)$){};
	\node[G] (v1) at ($(ha1)+(72:6pt)$){};	
	\node[G] (v2) at ($(ha2)+(144:6pt)$){};
	\node[G] (v3) at ($(ha3)+(-144:6pt)$){};
	\node[G] (v4) at ($(ha4)+(-72:6pt)$){};	
	\node[G] (v5) at ($(ha0)-(0:6pt)$){};
	\node[G] (v6) at ($(ha1)-(72:6pt)$){};	
	\node[G] (v7) at ($(ha2)-(144:6pt)$){};
	\node[G] (v8) at ($(ha3)-(-144:6pt)$){};
	\node[G] (v9) at ($(ha4)-(-72:6pt)$){};
	\draw[Ge] (v0)--(v1)--(v2)--(v3)--(v4)--(v5)--(v6)--(v7)--(v8)--(v9)--(v0);
\end{scope}

	\dtenprimgraph;
	\node[G] (va0) at (ha0) {};
	\node[G] (va12) at ($(ha1)+(120:5pt)$) {};
	\node[G] (va13) at ($(ha1)+(-120:5pt)$) {};	
	\node[G] (va2) at (ha2) {};
	\node[G] (va3) at (ha3) {};	
	\draw[Ge] (va0)--(va12)--(va2)--(va3)--(va13)--(va0);	
	\node[G] (vb0) at (hb0) {};
	\node[G] (vb12) at ($(hb1)+(-120:5pt)$) {};
	\node[G] (vb13) at ($(hb1)+(120:5pt)$) {};	
	\node[G] (vb2) at (hb2) {};
	\node[G] (vb3) at (hb3) {};	
	\draw[Ge] (vb0)--(vb12)--(vb2)--(vb3)--(vb13)--(vb0);	
	\draw[Ge] (va0)--(vb0);
	
\begin{scope}[shift={(8,0)}]
	\dtenprimgraph;
	\node[G] (va0) at ($(ha1)+(60:5pt)$) {};
	\node[G] (va1) at ($(ha2)+(120:5pt)$) {};
	\node[G] (va2) at (ha3) {};
	\node[G] (va3) at ($(ha1)-(60:5pt)$) {};
	\node[G] (va4) at ($(ha2)-(120:5pt)$) {};					
	\draw[Ge] (va0)--(va1)--(va2)--(va3)--(va4)--(va0);
	\node[G] (va5) at (ha0) {};
	\node[G] (vb5) at (hb0) {};		
	\node[G] (vb0) at ($(hb1)-(60:5pt)$) {};
	\node[G] (vb1) at ($(hb2)-(120:5pt)$) {};
	\node[G] (vb2) at (hb3) {};
	\node[G] (vb3) at ($(hb1)+(60:5pt)$) {};
	\node[G] (vb4) at ($(hb2)+(120:5pt)$) {};					
	\draw[Ge] (vb0)--(vb1)--(vb2)--(vb3)--(vb4)--(vb0);	
	\draw[Ge] (va0)--(va5)--(vb5)--(vb0);
\end{scope}	
\end{tikzpicture}
	\vspace*{-1.5em}
	\caption{Left: walking along the 10 edges of the thin black graph gives a tight walk containing all vertices, so no reconfiguration step is possible.
	Middle: walking along one cycle, the bridge, the second cycle, and then back along the bridge, gives a tight closed walk containing all vertices, so no reconfiguration step is possible.
	Right: no closed walk is tight, but the 4 middle vertices are frozen.}
\label{fig:frozen}	
\end{figure}

This generalizes the characterization of frozen vertices in the case of $H=K_3$ from~\cite{CerecedaHJ11}.
In general, frozen vertices can also arise in other situations, see Figure~\ref{fig:frozen} for an example, but these will not be relevant to the characterization theorem.

Finding any frozen vertex $v$ means $S(v)=\varepsilon$ for any solution sequence $S$.
This allows us to limit potentially realizable walks $Q$ to a single one, since even if our arbitrarily chosen vertex $q$ is not frozen, Lemma~\ref{lem:onePathImpliesAll} gives
	$Q = \red{S(q)} = \red{\alpha(W)}^{-1} \cdot  \red{S(v)} \cdot \red{\beta(W)} = \red{\alpha(W)}^{-1} \cdot \red{\beta(W)}$, for any walk $W$ from $v$ to $q$.
That is, we have the following necessary condition for a walk to be realizable:

\begin{corollary}
\label{cor:tightDeterminesQ}
	Let $\alpha,\beta$ be two $H$-colorings of $G$ and let $q$ be a vertex of $G$.
	If~$Q\in \pi(H)$ is realizable for $\alpha,\beta,q$,
	then for any $\alpha(C)$-tight closed walk in $H$, any vertex $v$ on $C$ and any walk $W$ from $v$ to $q$, we have 
	$Q = \red{\alpha(W)}^{-1} \cdot \red{\beta(W)}$.
\end{corollary}

Finally, we show how to find $\alpha$-tight closed walks by exploring walks $W$ such that $\alpha(W)$ is reduced. 

\begin{lemma}
\label{lem:tightAlgo}
There is an algorithm that given $G,H,\alpha$, finds an $\alpha$-tight walk or concludes there is none in time $\Oh(|V(G)|\cdot |E(G)|)$.
\end{lemma}
\begin{proof}
Consider the following directed graph $D$: its vertices are oriented edges of $G$ and there is an arc from $e$ to $e'$ when endpoints match ($\tau(e)=\iota(e')$) and $\alpha(e)\neq \alpha(e')^{-1}$.
Then directed cycles in $D$ are $\alpha$-tight closed walks in $G$, and conversely, any $\alpha$-tight closed walk in $G$ gives a directed cycle in $D$ (if some oriented edge of the closed walk is repeated, use only the fragment between the closest two repetitions).

$D$ has $2|E(G)|$ vertices and $\sum_{v\in V(G)} 2\binom{\deg(v)}{2} \leq |V(G)| \cdot \sum_{v \in V(G)} \deg(v) = \Oh(|V(G)|\cdot |E(G)|)$ arcs, so a directed cycle in $D$ can be found by depth-first search in time $\Oh(|V(G)|\cdot |E(G)|)$.
\end{proof}
We note that the prefix tree of walks $W$ such that $\alpha(W)$ is reduced gives a generalization of the layer construction~of~\cite{CerecedaHJ11} (infinite paths in it are walks that must contain the same oriented edge twice, so the fragment between repetitions is an $\alpha$-tight closed walk; conversely, any $\alpha$-tight closed walk implies an infinite walk $W$ such that $\alpha(W)$ is reduced, so an infinite path in the tree). 

\medskip

\section{Characterization of realizable walks}\label{sec:char}
In this section we prove the characterization theorem: the three necessary conditions described in the previous sections are enough to characterize all possible solutions to an \problem{$H$-Recoloring} instance, see Figure~\ref{fig:characterization}.
This is very unexpected, as it shows we can view the graphs as purely topological structures and the only remaining conditions to remember are a simple parity condition and the condition that tight closed walks are frozen---the fact that edges are actually discrete and cannot be stretched arbitrarily turns out to imply no further obstructions to reconfiguration (it only restricts the possible $H$-colorings, which are given on input).
The algorithm in Section~\ref{sec:algo} will use the theorem to find a concise description of the set of all realizable walks, in particular to find one such walk.

\begin{theorem}
\label{thm:characterization}
	Let $\alpha,\beta$ be two $H$-colorings of $G$.
	Consider any vertex $q$ of $G$ and let $Q\in\pi(H)$ be a reduced walk in $H$ from $\alpha(q)$ to $\beta(q)$.
	Then $Q$ is realizable for $\alpha,\beta,q$ if and only if
	\begin{itemize}
	\item
		$Q$ is topologically valid for $\alpha,\beta,q$,
	\item
		$Q$ has even length,
	\item
		for every $\alpha$-tight walk, any vertex $v$ on this walk and any walk $W$ from $v$ to $q$,\\
		$Q = \red{\alpha(W)}^{-1} \cdot \red{\beta(W)}$.
	\end{itemize}
	Furthermore, there is an $\Oh(|V(G)|^2+|V(G)|\cdot|Q|)$-time algorithm that given $G,H,\alpha,\beta$ and given a walk $Q$ satisfying these conditions,
	outputs a reconfiguration sequence (as a sequence of color changes) such that $S(q)=Q$, $S(v)$ is reduced and $|S(v)| \leq 2|V(G)| + |Q|$ for all $v\in V(G)$.
\end{theorem}

\begin{proof}
	If an $H$-recoloring sequence is given, then the conditions are satisfied by
	Corollary~\ref{cor:topoValid} and Corollary~\ref{cor:tightDeterminesQ}, which proves the `only if' half.
	
	Consider now a reduced walk $Q\in \pi(H)$ that satisfies the above conditions. For every vertex $v\in V(G)$ let $S_v = \red{\alpha(W)}^{-1} \cdot  Q \cdot \red{\beta(W)}$ for some walk $W$ from $q$ to $v$; by Lemma~\ref{lem:oneImpliesAll}, this does not depend on how $W$ is chosen.
	In particular $S_q = Q$.
	We will show an $H$-recoloring sequence $S$ from $\alpha$ to $\beta$ such that $S(v)=S_v$ for all $v\in V(G)$.
	The idea is that $S_v$ define a correct $H$-recoloring sequence for each edge, but it remains to order changes of different vertices into one reconfiguration sequence.
	Each edge gives a condition on who should recolor first
	and it turns out to be enough to respect these conditions.
	This is impossible if and only if there is a cycle of conditions,
	which turns out to be exactly an $\alpha$-tight cycle.
	\smallskip

	Formally, observe first that since $|Q|$ is even, $|\alpha(W)|=|W|=|\beta(W)|$ and reducing preserves parity, we have that each $S_v$ has even length.
	
\begin{figure}[H]
	\centering
	\begin{tikzpicture}[scale=0.9]
	\dninegraph;
	\node[G] (v0) at (ha0) {};
	\node[G,label=above:$v_3$] (v1) at (ha1) {};
	\node[G,label=left:$v_2$] (v2) at (ha2) {};
	\node[G,label=left:$v_1$] (v3) at ($(ha3)+(180:4pt)$) {};
	\node[G1,label=left:$q$] (v4) at ($(ha4)+(-120:4pt)$) {};
	\node[G] (v5) at ($(ha3)-(180:4pt)$) {};		
	\node[G] (v6) at ($(ha4)-(-120:4pt)$) {};	
	\node[G] (v7) at (ha5) {};
	\draw[Ge] (v0)--(v1)--(v2)--(v3)--(v4)--(v5)--(v6)--(v7)--(v0);	
	\node at (1.5,-1.5) {\Large$\alpha$};

\begin{scope}[shift={(5.5,0)}]
	\dninegraph;
	\node[G,label=60:$v_3$] (v0) at (ha0) {};
	\node[G,label=90:$v_2$] (v1) at (ha1) {};
	\node[G,label=90:$v_1$] (v2) at ($(ha2)+(120:4pt)$) {};
	\node[G1,label=left:$q$] (v3) at ($(ha3)+(180:4pt)$) {};
	\node[G] (v4) at ($(ha2)-(120:4pt)$) {};
	\node[G] (v5) at ($(ha3)-(180:4pt)$) {};		
	\node[G] (v6) at (ha4) {};	
	\node[G] (v7) at (ha5) {};
	\draw[Ge] (v0)--(v1)--(v2)--(v3)--(v4)--(v5)--(v6)--(v7)--(v0);	
	\node at (1.5,-1.5) {\Large$\beta$};	
\end{scope}	

\begin{scope}[shift={(11,0)}]
	\dninegraph;	
	\node[G1] (r0) at (ha4) {};
	\node[G1] (r1) at (ha3) {};	
	\draw[Gve,thick,red,->] (r0) to (r1);
	\node[Gv] (g1) at ($(ha5)-(-60:5pt)$) {};
	\node[Gv] (g2) at ($(ha0)-(0:5pt)$) {};	
	\node[Gv] (g3) at ($(ha1)-(60:5pt)$) {};
	\node[Gv] (g4) at ($(ha2)-(120:5pt)$) {};		
	\draw[Gve,thick,green!80!black,->] (r0) to (g1) (g1) to (g2) (g2) to (g3) (g3) to (g4) (g4) to (r1); 	
	\node[Gv] (b1) at ($(ha5)+(-60:4pt)$) {};
	\node[Gv] (b2) at ($(ha0)-(0,4pt)$) {};	
	\node[Gv] (b3) at ($(hb0)-(0,4pt)$) {};	
	\node[Gv] (b4) at (hb2) {};	
	\node[Gv] (b5) at (hb1) {};	
	\node[Gv] (b6) at ($(hb0)+(0,4pt)$) {};				
	\node[Gv] (b7) at ($(ha0)+(0,4pt)$) {};		
	\node[Gv] (b8) at ($(ha1)+(60:4pt)$) {};
	\node[Gv] (b9) at ($(ha2)+(120:4pt)$) {};		
	\draw[Gve,thick,blue!80!black,->] (r0) to (b1) (b1) to (b2) (b2) to (b3) (b3) to (b4) (b4) to (b5) (b5) to (b6) (b6) to (b7) (b7) to (b8) (b8) to (b9) (b9) to (r1); 	
	\node at (1.5,-1.5) {\large$S(q)?$};	
\end{scope}	
\end{tikzpicture}
	\vspace*{-2em}	
	\caption{Two $H$-colorings $\alpha$, $\beta$ of an 8-cycle, where $H$ is the gray graph on 9 vertices. Even though no vertex is frozen, $\alpha$ cannot be reconfigured to $\beta$.
	The red and green walks are not realizable for $\alpha,\beta,q$ because of parity.
	The blue walk has good parity, but is not topologically valid (imagine continuously deforming the 8-cycle by pulling $q$ along the blue path---the cycle would necessarily end up stretched around the triangle).}
\label{fig:characterization}
\end{figure}
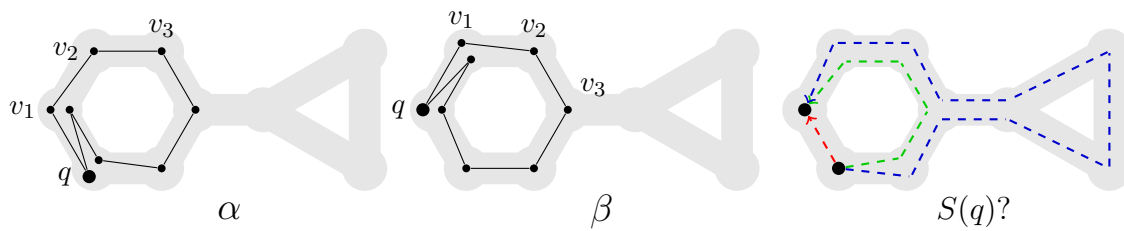

\pagebreak

	Consider two adjacent vertices $u,v\in V(G)$.
	Let $W$ be any walk from $q$ to $u$.
	Then $W$ followed by $(u,v)$ is a walk from $q$ to $v$, hence by definition $S_v =$
	$$= \red{\alpha(W (u,v))}^{-1} \cdot  Q \cdot \red{\beta(W (u,v))} =$$
	$$= (\alpha(v),\alpha(u)) \cdot \red{\alpha(W)}^{-1} \cdot  Q \cdot  \red{\beta(W)} \cdot (\beta(u),\beta(v)) =$$
	$$= (\alpha(v),\alpha(u)) \cdot S_u \cdot (\beta(u),\beta(v)).$$
	\smallskip

	Let
	$$S_u = (a_0,a_1)(a_1,a_2)(a_2,a_3)\dots (a_{n-1},a_n),$$
	$$S_v = (b_0,b_1)(b_1,b_2)(b_2,b_3)\dots (b_{m-1},b_m).$$
	Note that using concatenation (as opposed to $\cdot$) we write the exact sequence of edges of a walk.
	In particular $S_u$ is by definition reduced, so $a_i\neq a_{i+2}$, similarly for $S_v$.
	Now $\alpha(u)=a_0, \alpha(v)=b_0$ and $\beta(u)=a_n, \beta(v)=b_m$, hence
	$$S_v = (b_0,a_0) \cdot S_u \cdot (a_n,b_m).$$
	If $S_u$ is non-empty, $n \geq 1$, then by the parity condition $n\geq 2$ and
	$$S_v = \red{(b_0,a_0)(a_0,a_1)}(a_1,a_2)\dots (a_{n-2},a_{n-1}) \red{(a_{n-1},a_n)(a_n,b_m)}.$$
	There are two cases, depending on whether $\red{(b_0,a_0)(a_0,a_1)}$ cancels out to $\varepsilon$.
	Either it does, that is $(b_0,a_0) = (a_0,a_1)^{-1}$, in which case
	$$S_v = (a_1,a_2)(a_2,a_3)\dots (a_{n-2},a_{n-1}) \red{(a_{n-1},a_n)(a_n,b_m)}$$
	which means $b_0=a_1$, $b_1=a_2$, \dots, $b_{n-2}=a_{n-1}$ 	\hfill\textbf{(case $\boldsymbol{u\rightarrow v}$)}\\
	or it does not, in which case
	$$S_v = (b_0,a_0)(a_0,a_1)(a_1,a_2)\dots (a_{n-2},a_{n-1}) \red{(a_{n-1},a_n)(a_n,b_m)}$$	
	which means $b_1=a_0$, $b_2=a_1$, \dots, $b_n=a_{n-1}$	\hfill\textbf{(case $\boldsymbol{u\leftarrow v}$)}.
	\bigskip
	\pagebreak[3]	

	For two adjacent vertices $u,v\in V(G)$ such that $S_u$ and $S_v$ are non-empty,
	let us write $u\rightarrow v$ in the first case and $u \leftarrow v$ in the other, as defined above.
	We have $u\rightarrow v$ iff $v \leftarrow u$, otherwise $b_0=a_1=b_2$ (if $u\rightarrow v$ and $v\rightarrow u$) or $a_0=b_1=a_2$ (if $u\leftarrow v$ and $v\leftarrow u$), contradicting that $S_u,S_v$ are reduced walks.
	
	Furthermore, the $\rightarrow$ relation has no cycles.
	Suppose to the contrary that there exist $v_0$, $v_1$, \dots, $v_{l-1},v_0$ ($l\geq 3$)
	such that $v_i \rightarrow v_{i+1}$ for $i\in \mathbb{Z}_l$.
	We will write $S_v^j$ for the $j$-th vertex of $S_v$.
	Then this is an $\alpha$-tight walk:
	indeed, arrows imply adjacency in $G$, and
	$v_i \rightarrow v_{i+1} \rightarrow v_{i+2}$ implies that 
	$$\alpha(v_{i+2}) = S_{v_{i+2}}^0 = S_{v_{i+1}}^1 = S_{v_i}^2 \neq S_{v_i}^0 = \alpha(v_i).$$
	Therefore by the last condition we have $Q = \red{\alpha(W)}^{-1} \cdot \red{\beta(W)}$ and for any walk $W$ from $v_i$ to $q$,
	$$S_{v_i} =
	\red{\alpha(W)} \cdot Q \cdot \red{\beta(W)}^{-1} =
	\red{\alpha(W)} \cdot \red{\alpha(W)}^{-1} \cdot \red{\beta(W)} \cdot \red{\beta(W)}^{-1} =
	\varepsilon.$$
	But we did not assign arrows between vertices whose sequences are empty, a contradiction.

	Therefore there is an ordering $v_1,v_2,\dots,v_{|V(G)|}$ of $V(G)$ such that if $v_i \rightarrow v_j$ then $i<j$.
	We claim the following is a valid $H$-recoloring sequence from $\alpha$ to $\beta$.
	Recolor:\\
	$v_1$ from $S_{v_1}^0$ to $S_{v_1}^2$\quad,\quad
	$v_2$ from $S_{v_2}^0$ to $S_{v_2}^2$\quad,\quad
	\dots\quad,\quad
	$v_n$ from $S_{v_n}^0$ to $S_{v_n}^2$,\\
	$v_1$ from $S_{v_1}^2$ to $S_{v_1}^4$\quad,\quad
	$v_2$ from $S_{v_2}^2$ to $S_{v_2}^4$\quad,\quad
	\dots\quad,\quad
	$v_n$ from $S_{v_n}^2$ to $S_{v_n}^4$,\\
	$v_1$ from $S_{v_1}^4$ to $S_{v_1}^6$\quad , \quad \dots \quad \dots\\
	We continue in this order (disregarding any undefined recolorings to $S^j_{v_i}$ for $j>|S_{v_i}|$).
	Because of the parity condition, every vertex $v_i$ eventually gets recolored to the last color in $S_{v_i}$, which is $\beta(v_i)$;
	that is, the coloring we reach is indeed $\beta$.
	
	To check that it is a valid $H$-recoloring sequence, consider any edge $\{u,v\}$ of $G$ and define $a_i,b_i$ as above.
	If both $S_u$ and $S_v$ are empty, then $\{u,v\}$ gets constantly mapped to the same edge $\{\alpha(u),\alpha(v)\}$ of $H$.
	If exactly one of $S_u, S_v$ is empty, say $S_u$, then $S_v=(b_0,a_0)\cdot S_u \cdot (a_0, b_m) = (b_0,a_0)(a_0,b_m)$ where $b_0\neq b_m$ (and $m=2$).
	Thus $b_1=a_0$, so $\{u,v\}$ gets mapped initially to $\{\alpha(u),\alpha(v)\}=\{a_0,b_0\}$ and then to $\{a_0,b_2\}=\{b_1,b_2\}$,
	which is an edge of $H$.
	If both $S_u$ and $S_v$ are non-empty, then assume without loss of generality $u \rightarrow v$ (otherwise swap $u$ and $v$).
	We have
	\begin{alignat*}{4}
		S_u &= (a_0,a_1)&& (a_1,a_2)(a_2,a_3)\dots (a_{n-2},a_{n-1})(a_{n-1},a_n),\\
		S_v &=          && (a_1,a_2)(a_2,a_3)\dots (a_{n-2},a_{n-1}) \red{(a_{n-1},a_n)(a_n,b_m)}
	\end{alignat*}
	Thus $(u,v)$ gets mapped initially to $(\alpha(u),\alpha(v))=(a_0,b_0)=(a_0,a_1)$
	and then to $(a_1,a_2)$, $(a_3,a_2)$, $(a_3,a_4)$, \dots, ending in either $(a_n,a_{n-1})$ or $(a_n,a_{n+1})$ depending on whether $\red{(a_{n-1},a_n)(a_n,b_m)}=\varepsilon$.
	This is again always an edge of $H$ (because $S_u, S_v$ are walks in $H$). Thus the $H$-coloring condition is never violated on any edge and the sequence is a valid $H$-recoloring sequence.
	
	The algorithm first checks that $|Q|$ is even. Then, it has to choose some walks to define $S_v$ for $v \in V(G)$; choosing a shortest paths from $q$ (in time $\Oh(|E(G)|)$) guarantees $|S_v| \leq 2|V(G)|+|Q|$.
	Then, for each edge $uv$ of $G$, it checks whether $u \to v$ or $v\to u$ holds, by inspecting the first edges of $S_v$ and $S_u$ in constant time, $\Oh(|E(G)|)$ in total.
	The ordering $v_1,\dots,v_{|V(G)|}$ (a topological ordering of the arrow graph) is constructed in $\Oh(|E(G)|)$ time (if none is found, we can output a tight closed walk, in fact a tight cycle).
	Finally it outputs the sequence of color changes given by $S_v$ in the above order, in time linear in the total number of color changes, which is $\Oh(|V(G)| \cdot(|V(G)|+|Q|))$.
	The algorithm can check whether the conditions on $Q$ were really satisfied by checking the consecutive colors on each edge as in the previous paragraph, in total time $\Oh(|E(G)| \cdot(|V(G)|+|Q|))$; if at some point the check fails, this is a contradiction, which means that $Q$ could not have been topologically valid.
	If we wanted to output the entire $H$-coloring at each step, this makes the output $|V(G)|$ times larger, requiring $\Oh(|V(G)|^2 \cdot(|V(G)|+|Q|))$ total time.
\end{proof}

The running time does not depend on $H$ at all, because we only inspect images of some edges in $G$; $H$ could indirectly cause $Q$ to be long, but we will construct realizable walks $Q$ of polynomial length.

As the proof of the characterization theorem produces a solution sequence where all vertex walks are reduced, any sequence where this is not true can be shortened.

\begin{corollary}
\label{cor:shortestIsNonbacktracking}
	Let $\alpha,\beta$ be two $H$-colorings of $G$.
	Let $S=\sigma_0,\dots,\sigma_l$ be an $H$-recoloring sequence between $\sigma_0=\alpha$ and $\sigma_l=\beta$ such that $l$ is minimized.
	Then for each vertex $v$ of $G$, $S(v)$ is reduced.
\end{corollary}
\begin{proof}
	Suppose $S(q)$ is not reduced for some $q$.
	Let $Q=\red{S(q)}$.
	By the above theorem we know from one side that $Q$ is realizable.
	From the other side we obtain a solution sequence $S'$ such that $S'(v)=\red{S'(v)}$ for all $v$ and $\red{S'(q)}=Q=\red{S(q)}$.
	By Lemma~\ref{lem:onePathImpliesAll},
	this implies $S'(v)=\red{S(v)}$, for all $v$.
	But $\red{S(v)}$ is always no longer than $S(v)$, and $\red{S(q)}$ is strictly shorter than $S(q)$.
	Since the number of recoloring steps is equal to half the sum of lengths of all $S(v)$, $S$ was not shortest.
\end{proof}

\section{Calculations in the fundamental groupoid}\label{sec:calculations}
The goal of this section is to prove Lemma~\ref{lem:topoValidEnumeration}, which describes algorithmically the set of topologically valid walks.
This follows from well-known calculations in the fundamental groupoid of graphs, which we recall here.

Any algorithm will need to limit the number of closed walks considered.
The standard way to do that is as follows: fix a vertex $q\in V(G)$ and
an arbitrary spanning tree $T$ of $G$ (a minimal connected subgraph that includes all vertices).
For $e\in E(G)\setminus E(T)$ and an arbitrarily fixed orientation $(\iota(e),\tau(e))$ of $e$ define the \emph{fundamental cycle} $C_e$ as the closed walk that goes from $q$ to $\iota(e)$ along the unique path that connects them in $T$, then to $\tau(e)$ through $e$, then back to $q$ along the unique path in $T$. 
There are $|E(G)| - |E(T)| = |E(G)|-|V(G)|+1$ fundamental cycles and together they generate all other cycles (see for example Lemma 1.2. in \cite{kwak2007graphs}):

\begin{fact}
\label{fact:fundCycle}
	Let $C$ be any closed walk from $q$ to $q$ in $G$.
	Then $\red{C} = \red{C_{e_1}}^{s_1} \cdot\ \dots\ \cdot \red{C_{e_n}}^{s_n}$ where  $e_1,\dots,e_n \in E(G)\setminus E(T)$ are the consecutive non-tree edges of $C$ and $s_i\in\{-1,+1\}$ are chosen to match their orientation.
\end{fact}

This allows to limit the number of conjugacy equations defining topological validity to polynomially many ($|E(G)\setminus E(T)|$ to be exact). It is also folklore that  conjugacy equations can be solved in polynomial time: 

\begin{fact}
\label{fact:topoEquivAlgo}
	Given two $H$-colorings $\alpha,\beta$ of a graph $G$ and a vertex $q$, one can find in time $\Oh(|E(G)|\cdot|V(G)|+|E(H)|)$ a walk $Q\in\pi(H)$ that is topologically valid for $\alpha,\beta,q$, or conclude there is none.
\end{fact}
\begin{proof}
	By definition, $Q$ is topologically valid if and only if for every closed walk $C$ from $q$ to $q$ we have	$\red{\beta(C)} = Q^{-1} \cdot \red{\alpha(C)} \cdot Q$.
	By Fact~\ref{fact:fundCycle}, this is equivalent to satisfying the equation for each fundamental cycle $C_e$.
	Let $W$ be any walk from $\beta(q)$ to $\alpha(q)$ in $H$.
	Then $Q$ satisfies the equations if and only if $Q \cdot W$ satisfies $W \cdot \red{\beta(C_e)} \cdot W^{-1} = (Q \cdot W)^{-1} \cdot \red{\alpha(C_e)} \cdot (Q \cdot W)$ for each fundamental cycle $C_e$.
	In this form, we have polynomially many equations where each of the walks $Q \cdot W$, $\red{\alpha(C_e)}$ and $W \cdot \red{\beta(C_e)} \cdot W^{-1}$ is a closed walk from $\alpha(q)$ to $\alpha(q)$ in $H$.

	Denote by $\pi(H,\alpha(q))$ the subset of $\pi(H)$ given by closed walks from $\alpha(q)$ to $\alpha(q)$. It is easy to check that $\pi(H,\alpha(q))$ is a group (under $\cdot$);
	moreover, it is the free group generated by the fundamental cycles of $H$ as described in Fact~\ref{fact:fundCycle} (see for example Lemma 1.1., 1.2. in \cite{kwak2007graphs}).
	Finding an element $Q\cdot W$ satisfying the above equations in $\pi(H,\alpha(q))$ is therefore the Simultaneous Conjugacy Search Problem in a free group, for which a linear time algorithm is described in Theorem 6.5. of~\cite{MyasnikovU08}.
	The size of the input to this algorithm can be bounded by the number of equations $|E(G)\setminus E(T)|$ times the length of $\red{\alpha(C_e)}$ and $W \cdot \red{\beta(C_e)} \cdot W^{-1}$ in each equation, which is $\Oh(|V(G)|)$.
	Additionally, we need to compute a spanning tree of $H$ and give the edges outside of it, in $\Oh(|E(H)|)$ time, to present $\pi(H,\alpha(q))$ as a free group.
\end{proof}

To describe all valid walks we will need the following.
For a non-empty closed walk $C\in \pi(H)$ we define the \emph{primitive root} of $C$ as the unique $R\in \pi(H)$ such that $C=R^n$ for some $n\in\mathbb{N}$ such that $n$ is maximized.
Note that if $R$ is a primitive root, then the primitive root of $W \cdot R^n \cdot W^{-1}$ is $W \cdot R \cdot W^{-1}$ (for $n\geq 1$ and $W \in \pi(H)$ such that $W \cdot R$ is defined), for example. 
It is a routine exercise to check the primitive root is well defined, can be computed in linear time, and that the following holds (see e.g. Lemma 2.1. of \cite{madlener1980string}):

\begin{fact}
\label{fact:commuteRoot}
	Let $C_1,C_2 \in \pi(H)$.
	Then $C_1$ and $C_2$ commute, i.e., $C_1\cdot C_2=C_2\cdot C_1$, if and only if
	$C_1=\varepsilon$ or $C_2=\varepsilon$ or both have the same primitive root or one root is the inverse of the other. 
\end{fact}

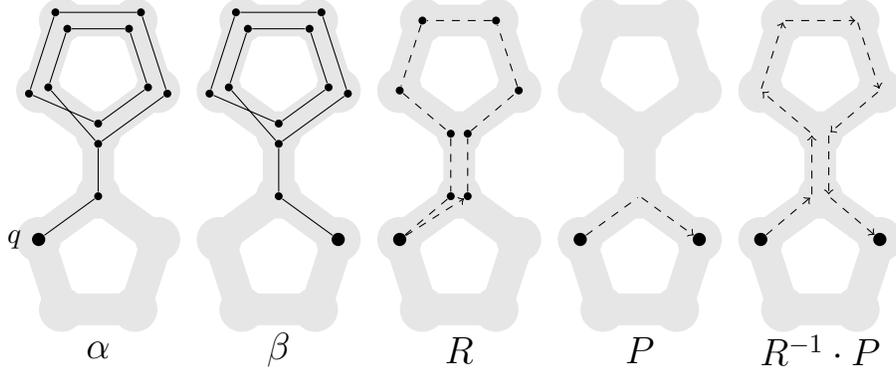
\begin{figure}[H]
	\centering
	\begin{tikzpicture}[scale=0.8]
\begin{scope}[shift={(0,0)},scale=0.8,rotate=-90]
	\dtengraph;
	\node[G] (v0) at ($(ha0)+(0:6pt)$){};
	\node[G] (v1) at ($(ha1)+(72:6pt)$){};	
	\node[G] (v2) at ($(ha2)+(144:6pt)$){};
	\node[G] (v3) at ($(ha3)+(-144:6pt)$){};
	\node[G] (v4) at ($(ha4)+(-72:6pt)$){};	
	\node[G] (v5) at ($(ha0)-(0:6pt)$){};
	\node[G] (v6) at ($(ha1)-(72:6pt)$){};	
	\node[G] (v7) at ($(ha2)-(144:6pt)$){};
	\node[G] (v8) at ($(ha3)-(-144:6pt)$){};
	\node[G] (v9) at ($(ha4)-(-72:6pt)$){};
	\draw[Ge] (v0)--(v1)--(v2)--(v3)--(v4)--(v5)--(v6)--(v7)--(v8)--(v9)--(v0);
	\node[G] (v10) at (hb0) {};
	\node[G1,label=left:$q$] (v11) at (hb1) {};
	\draw[Ge] (v0)--(v10)--(v11);	
	\node at (4.5,0) {\Large$\alpha$};
\end{scope}

\begin{scope}[shift={(3,0)},scale=0.8,rotate=-90]
	\dtengraph;
	\node[G] (v0) at ($(ha0)+(0:6pt)$){};
	\node[G] (v1) at ($(ha1)+(72:6pt)$){};	
	\node[G] (v2) at ($(ha2)+(144:6pt)$){};
	\node[G] (v3) at ($(ha3)+(-144:6pt)$){};
	\node[G] (v4) at ($(ha4)+(-72:6pt)$){};	
	\node[G] (v5) at ($(ha0)-(0:6pt)$){};
	\node[G] (v6) at ($(ha1)-(72:6pt)$){};	
	\node[G] (v7) at ($(ha2)-(144:6pt)$){};
	\node[G] (v8) at ($(ha3)-(-144:6pt)$){};
	\node[G] (v9) at ($(ha4)-(-72:6pt)$){};
	\draw[Ge] (v0)--(v1)--(v2)--(v3)--(v4)--(v5)--(v6)--(v7)--(v8)--(v9)--(v0);
	\node[G] (v10) at (hb0) {};
	\node[G1] (v11) at (hb4) {};
	\draw[Ge] (v0)--(v10)--(v11);
	\node at (4.5,0) {\Large$\beta$};		
\end{scope}

\begin{scope}[shift={(6,0)},scale=0.8,rotate=-90]
	\dtengraph;
	\node[G] (v0) at ($(ha0)+(90:5pt)$){};
	\node[G] (v1) at (ha1){};	
	\node[G] (v2) at (ha2){};
	\node[G] (v3) at (ha3){};
	\node[G] (v4) at (ha4){};	
	\node[G] (v5) at ($(ha0)-(90:5pt)$){};	
	\node[G] (v6) at ($(hb0)-(90:5pt)$) {};
	\node[G1] (v7) at (hb1) {};
	\node[G] (v8) at ($(hb0)+(90:5pt)$) {};	
	\draw[Gve,->] (v8)--(v0)--(v1)--(v2)--(v3)--(v4)--(v5)--(v6)--(v7)--(v8);
	\node at (4.5,0) {\Large$R$};	
\end{scope}

\begin{scope}[shift={(9,0)},scale=0.8,rotate=-90]
	\dtengraph;
	\node[G1] (v0) at (hb1){};
	\node[Gv] (v1) at (hb0){};	
	\node[G1] (v2) at (hb4){};
	\draw[Gve,->] (v0) to (v1) (v1) to (v2);
	\node at (4.5,0) {\Large$P$};	
\end{scope}

\begin{scope}[shift={(12,0)},scale=0.8,rotate=-90]
	\dtengraph;
	\node[G1] (v0) at (hb4) {};	
	\node[Gv] (v1) at ($(hb0)+(90:5pt)$) {};		
	\node[Gv] (v2) at ($(ha0)+(90:5pt)$){};
	\node[Gv] (v3) at (ha1){};	
	\node[Gv] (v4) at (ha2){};
	\node[Gv] (v5) at (ha3){};
	\node[Gv] (v6) at (ha4){};	
	\node[Gv] (v7) at ($(ha0)-(90:5pt)$){};	
	\node[Gv] (v8) at ($(hb0)-(90:5pt)$) {};
	\node[G1] (v9) at (hb1) {};
	\foreach \x [remember=\x as \lastx (initially 0)] in {1,...,9}
	{ \draw[Gve,->] (v\x) to (v\lastx); }
	\node at (4.5,0) {\Large$R^{-1}\cdot P$};	
\end{scope}
\end{tikzpicture}
	\caption{In this example, let $C$ be the shortest non-trivial walk from $q$ to $q$ in the thin black graph $G$.  Its image $\alpha(C)$ and $\beta(C)$ winds twice around the root $R$ ($\red{\alpha(C)} = R^2$).
	The topologically valid paths are exactly $\{R^n \cdot P \mid n\in\mathbb{Z}\}$ (think about deforming $\alpha$ by pulling $q$: one can pull it once or more around the top cycle by rotating all of $\alpha$, but this is impossible for the bottom cycle if we want to end at $\beta$).}
\label{fig:valid}
\end{figure}

We now show that whenever a cycle $C$ maps to a non-trivial cycle in $H$, possible solution sequences can only differ in the number of times they wind around this cycle; see Figure~\ref{fig:valid}.

\begin{lemma}
\label{lem:cycleTopoValid}
	Let $\alpha,\beta$ be two $H$-colorings of $G$ and let $q$ be a vertex of $G$.
	Let $P\in\pi(H)$ be topologically valid for $\alpha,\beta,q$.
	Then $Q\in\pi(H)$ is topologically valid for $\alpha,\beta,q$
	if and only if
	for every closed walk $C$ in $G$ starting and ending in $q$ such that $\red{\alpha(C)}\neq\varepsilon$, we have
	$$Q = R^n \cdot P$$
	for some $n\in\mathbb{Z}$,
	where $R\in\pi(H)$ is the primitive root of $\red{\alpha(C)}$.
\end{lemma}
\begin{proof}
	Suppose $Q$ is topologically valid and let $C$ be a closed walk starting and ending in $q$.
	Then by definition we have $\red{\beta(C)} = P^{-1} \cdot \red{\alpha(C)} \cdot P$ and $\red{\beta(C)} = Q^{-1} \cdot \red{\alpha(C)} \cdot Q$.
	Therefore	
	$$ P^{-1} \cdot \red{\alpha(C)} \cdot P =
	 Q^{-1} \cdot \red{\alpha(C)} \cdot Q$$
	$$ Q \cdot P^{-1} \cdot \red{\alpha(C)} =
	 \red{\alpha(C)} \cdot Q \cdot P^{-1}$$	 
	So $\red{\alpha(C)}$ commutes with $Q \cdot P^{-1}$.
	Therefore, if $\red{\alpha(C)}\neq\varepsilon$ and $R$ is the primitive root of $\red{\alpha(C)}$,
	then $Q\cdot P^{-1}=R^n$ for some $n\in\mathbb{Z}$ by Fact~\ref{fact:commuteRoot}.
	\smallskip
	
	For the other side, suppose that for every closed walk $C$ from $q$ to $q$ with $\red{\alpha(C)}\neq \varepsilon$ and a primitive root $R$ of $\red{\alpha(C)}$,
	there is an $n\in\mathbb{Z}$ such that $Q = R^n \cdot P$.
	Then for every closed walk $C$ from $q$ to $q$, $\red{\alpha(C)}$ commutes with $Q \cdot P^{-1}$, because either $\red{\alpha(C)}=\varepsilon$ or $\red{\alpha(C)}=R^k$ and $Q\cdot P^{-1} = R^n$ for some $R\in\pi(H)$ and $n,k\in\mathbb{Z}$.
	Thus $\red{\beta(C)} =$
	$$= P^{-1} \cdot \red{\alpha(C)} \cdot P =$$
	$$= Q^{-1} \cdot Q \cdot P^{-1} \cdot \red{\alpha(C)} \cdot P =$$
	$$= Q^{-1} \cdot \red{\alpha(C)} \cdot Q \cdot P^{-1} \cdot P =$$
	$$= Q^{-1} \cdot \red{\alpha(C)} \cdot Q,$$
	which shows the topological validity of $Q$.
\end{proof}

The above lemma allows us to describe the set of all topologically valid walks:

\pagebreak

\begin{lemma}
\label{lem:topoValidEnumeration}
	Let $\alpha,\beta$ be $H$-colorings of $G$ and $q$ a vertex of $G$.
	Consider the set $\Pi\subseteq\pi(H)$ of topologically valid walks for $\alpha,\beta,q$.
	One of the following holds:
	\begin{enumerate}[itemsep=3pt,topsep=3pt,partopsep=3pt,parsep=0pt]
		\item[0.] $\Pi = \emptyset.$
		\item[1.] $\Pi = \{Q\}$ for some $Q\in\pi(H)$.
		\item[2.] $\Pi = \{R^n \cdot P \mid n \in \mathbb{Z}\}$ for some $R,P\in\pi(H)$.
		\item[3.] $\Pi$ contains all reduced walks from $\alpha(q)$ to $\beta(q)$.
	\end{enumerate}
	Moreover, there is an algorithm that given $G,H,\alpha,\beta,q$ decides in time  $\Oh(|E(G)|\cdot|V(G)|+|E(H)|)$ which case holds and outputs $Q$ or $R,P$ in cases 1,2.
\end{lemma}
\begin{proof}
	Use Fact~\ref{fact:topoEquivAlgo} to compute a topologically valid walk $P\in\pi(H)$ for $\alpha,\beta,q$ in time $\Oh(|E(G)|\cdot|V(G)|+|E(H)|)$.
	If there is none, we immediately answer \emph{case 0}.
	Fix an arbitrary spanning tree, compute the elements $\red{\alpha(C_e)}$ for all fundamental cycles of $G$ (in total time $\Oh(|E(G)|\cdot|V(G)|)$) and if some for $e$ it is non-empty, check if it commutes with all other elements (again in time $\Oh(|E(G)|\cdot|V(G)|)$, since $|\red{\alpha(C_e)}|=\Oh(|V(G)|)$).
	One of the following holds:
	
	\begin{enumerate}[itemsep=4pt,topsep=4pt,partopsep=4pt,parsep=0pt]
	\item[a)] For every $C_e$, $\red{\alpha(C_e)}=\varepsilon$.
	Then by Fact~\ref{fact:fundCycle}, for every closed walk $C$ from $q$ to $q$, $\red{\alpha(C)}=\varepsilon$.
	By Lemma~\ref{lem:cycleTopoValid}, vacuously, every walk $Q\in\pi(H)$ from $\alpha(q)$ to $\beta(q)$ is topologically valid.
		
	\item[b)] There is a $C_e$  such that $\red{\alpha(C_e)}\neq \varepsilon$ and for every $C_f$, $\red{\alpha(C_f)}$ commutes with $\red{\alpha(C_e)}$.
	Then by Fact~\ref{fact:fundCycle}, for every closed walk $C$ starting and ending in $q$,
	$\red{\alpha(C)}$ commutes with $\red{\alpha(C_e)}$.
	Let $R$ be the primitive root of $\red{\alpha(C_e)}$. 
	$R$ (or its inverse) is also the primitive root of every non-empty $\red{\alpha(C)}$,
	so by Lemma~\ref{lem:cycleTopoValid}, $Q$ is topologically valid iff $Q= R^n \cdot P$ for some $n\in\mathbb{Z}$.

	\item[c)]
	 There are $C_e, C_f$ such that $\red{\alpha(C_e)}$ and $\red{\alpha(C_f)}$ do not commute.
	Then we show $\Pi=\{P\}$. 
	Clearly $\red{\alpha(C_i)} \neq \varepsilon$, so let $R_i$ be the primitive root of $\red{\alpha(C_i)}$ for $i\in\{e,f\}$.
	Suppose $Q\in\pi(H)$ is topologically valid.
	Then by Lemma~\ref{lem:cycleTopoValid},
	$Q=R_e^{n_e} \cdot P$ and $Q=R_f^{n_f} \cdot P$ for some $n_e,n_f\in\mathbb{Z}$.
	Thus $R_1^{n_1} = R_2^{n_2}$.
	If this element has a primitive root ($n_e,n_f\neq 0$), then it is equal to both $R_1$ and $R_2$, implying that $\red{\alpha(C_1)}$ and $\red{\alpha(C_2)}$ have the same primitive root, contradicting Fact~\ref{fact:commuteRoot}.
	Therefore $n_1=n_2=0$, so $Q$ must be equal to $P$.
	\end{enumerate}
	We output, respectively, \emph{case 3.}, \emph{case 2.} and $R,P$, or \emph{case 1.} and $P$.
\end{proof}

\section{The main algorithm}\label{sec:algo}
In this section we give the main algorithm, which returns a description of all solution sequences, in particular telling whether there is one. It follows directly from the algorithm for describing topologically valid walks in Lemma~\ref{lem:topoValidEnumeration} by simply checking the two other conditions of Theorem~\ref{thm:characterization}.

\begin{theorem}
\label{thm:realizableEnumeration}
	Let $\alpha,\beta$ be $H$-colorings of $G$ and $q$ a vertex of $G$.
	Consider the set $\Pi'\subseteq\pi(H)$ of realizable walks for $\alpha,\beta,q$.
	One of the following holds:
	\begin{enumerate}[itemsep=3pt,topsep=3pt,partopsep=3pt,parsep=0pt]
		\item[0.] $\Pi' = \emptyset.$
		\item[1.] $\Pi' = \{Q\}$ for some $Q\in\pi(H)$.
		\item[2.] $\Pi' = \{R^n \cdot P \mid n \in \mathbb{Z}\}$ for some $R,P\in\pi(H)$.
		\item[3.] $\Pi'$ contains all reduced walks of even length from $\alpha(q)$ to $\beta(q)$.
	\end{enumerate}
	Moreover, there is an algorithm that given $G,H,\alpha,\beta,q$ decides in time $\Oh(|E(G)|\cdot|V(G)|+|E(H)|)$ which case holds and outputs $Q$ or $R,P$ in cases 1,2.
\end{theorem}
\begin{proof}
	First, find any $\alpha$-tight closed walk and if there is one,
	let $Q$ be the only possibly realizable walk as in the last condition of Theorem~\ref{thm:characterization}.
	By running the algorithm from Theorem~\ref{thm:characterization} we can check whether it is indeed realizable and return either $\Pi'=\emptyset$ or $\Pi'=\{Q\}$.
	
	Assume now that there is no $\alpha$-tight walk.	
	Run the algorithm of Lemma~\ref{lem:topoValidEnumeration} to get a description of topologically valid walks $\Pi$ and consider the following cases:
	\begin{enumerate}
		\item[0.] $\Pi=\emptyset$. Then also $\Pi'=\emptyset$ (see Theorem~\ref{thm:characterization}).
		\item[1.] $\Pi=\{Q\}$ for some $Q\in\pi(H)$.
		Return $\Pi'=\{Q\}$ if $Q$ has even length and $\Pi'=\emptyset$ otherwise.
		\item[2.] $\Pi=\{R^n\cdot P \mid n\in\mathbb{Z}\}$ for some $R,P\in \pi(H)$.
		The only remaining condition is parity, so one of the following holds: 
		\begin{itemize}[itemsep=2pt,topsep=3pt,partopsep=3pt,parsep=0pt]
			\item $R$ is even and $P$ is odd: then $\Pi'=\emptyset$,
			\item $R$ is even and $P$ is even: then $\Pi'=\Pi$,
			\item $R$ is odd and $P$ is even: then $\Pi'= \{R^{2n}\cdot P \mid n\in\mathbb{Z}\}$,
			\item $R$ is odd and $P$ is odd: then $\Pi'= \{R^{2n}\cdot (R \cdot P) \mid n\in\mathbb{Z}\}$.
		\end{itemize}
		\item[3.] $\Pi$ contains all reduced walks from $\alpha(q)$ to $\beta(q)$.
		Then $\Pi'$ contains all reduced walks of even length from $\alpha(q)$ to $\beta(q)$.
		\qedhere
	\end{enumerate}
\end{proof}

The set of even walks from $\alpha(q)$ to $\beta(q)$ in $H$ is empty if and only if $H$ is bipartite and $\alpha(q),\beta(q)$ are on different sides of a bipartition. We can construct an even walk or conclude there is none in linear time. Thus in each case we decide whether there is a realizable walk $Q$ and if so, construct one of length bounded by the total running time, $\Oh(|E(G)|\cdot|V(G)|+|E(H)|)$. From $Q$, the algorithm of Theorem~\ref{thm:characterization} can compute an actual recoloring sequence (as a sequence of color changes) in time $\Oh(|V(G)|^2 + |V(G)|\cdot |Q|) = \Oh(|E(G)|\cdot|V(G)|^2+|E(H)|\cdot |V(G)|)$.
In particular, whenever some sequence exists, we output a sequence of polynomial length.
(Note that for non-square-free $H=K_4$, examples where shortest recoloring sequences have length exponential in $|V(G)|$ are known~\cite{BonsmaC09}).

\begin{corollary}
For square-free graphs $H$, \problem{$H$-Recoloring} can be decided in time $\Oh(|E(G)|\cdot|V(G)|+|E(H)|)$.
\end{corollary}

Shortest recoloring sequences can also be found in polynomial time with some more care. 

\begin{theorem}
For square-free graphs $H$, \problem{Shortest $H$-Recoloring} can be solved in time polynomial in the size of $G$ and $H$.
\end{theorem}
\begin{proof}
By Corollary~\ref{cor:shortestIsNonbacktracking}, it suffices to choose a walk $Q\in \Pi'$  from Theorem~\ref{thm:realizableEnumeration} minimizing
\begin{equation}\label{eqmin}
\sum_{v\in V(G)} \red{S(v)} = \sum_{v\in V(G)} |\red{\alpha(W_v)}^{-1} \cdot Q \cdot \red{\beta(W_v)}|,
\end{equation}
where $W_v$ is a walk from $q$ to $v$ (arbitrarily chosen).
In cases 0. and 1. this is trivial.
In case~2. ($Q=R^n \cdot P$, for any $n\in\mathbb{N}$) it is easy to see that $|n|\leq 2|V(G)|+ |P|$ in shortest sequences, since repeating $R$ will eventually lengthen all summands of~\eqref{eqmin}.
It thus suffices to compute~\eqref{eqmin} for all these possibilities for $n$.

In case 3., consider a realizable walk $Q$, i.e., any reduced walk of even length from $\alpha(q)$ to $\beta(q)$. Let $P_1$ be the longest common prefix of $Q$ and $\red{\alpha(W_v)}$, choosing $v\in V(G)$ to maximize its length. That is, $P_1$ is longest such that all of $P_1$ will reduce with $\red{\alpha(W_v)}^{-1}$ in  some summand of~\eqref{eqmin}.
Analogously, let $P_2$ bet the longest common suffix of $Q$ and some $\red{\beta(W_v)}^{-1}$.
Either $P_1$ and $P_2$ overlap, or $Q = P_1 Q' P_2$, for some $Q'\in \pi(H)$. In the latter case, since by definition no~element of $Q'$ will be reduced in any summand of \eqref{eqmin}, it can be written as
$$ \sum_{v\in V(G)} |\red{\alpha(W_v)}^{-1} \cdot Q \cdot \red{\beta(W_v)}|
= \sum_{v\in V(G)} \left( |\red{\alpha(W_v)}^{-1} \cdot P_1| + |Q'| + |P_2 \cdot \red{\beta(W_v)}| \right).$$
Thus we can guess $P_1$ by enumerating all prefixes of all $\red{\alpha(W_v)}$, similarly guess $P_2$ and guess how much they overlap. In case they do not overlap, the sum is minimized by taking $Q'$ to be an arbitrary shortest path of appropriate parity from the tail of $P_1$ to the head of $P_2$ in $H$.
Enumerating all possibilities for (the length of) $P_1$, $P_2$ and the overlap can be done in polynomial time, and a shortest path of given parity in $H$ can be found by duplicating every vertex, i.e., finding a shortest path in the tensor product $H \times K_2$.
\end{proof}

\section{Conclusions and future work}
\subsection*{The case $H=K_3$}
Our result generalizes the algorithm for \problem{$K_3$-Recoloring} of \cite{CerecedaHJ11} and recovers many of its features in a more general and perhaps more intuitive setting.
When limited to $H=K_3$ (a 3-cycle), there is only one possible root $R$ for closed walks in $H$ (and its inverse), they all commute.
Hence in the proof of Lemma~\ref{lem:topoValidEnumeration}, case 1. is impossible, while case 2. is the same as case 3.
So either no walk is topologically valid (that is, $\alpha,\beta$ are not homotopic), or all are.

This allows to simplify the algorithm for \problem{$K_3$-Recoloring}  substantially. Given an instance $\alpha,\beta$, if there is any solution sequence, any realizable walk, then we can find it as follows, knowing that all walks are topologically valid:
either there is some frozen vertex, which implies $Q_1=\varepsilon$ is realizable for this vertex, or no vertex is frozen, which implies that all even walks are realizable, in particular the walk $Q_2$ from $\alpha(q)$ to $\beta(q)$ of length 0 or 2.
In particular, we do not need to perform any of the calculations in Section~\ref{sec:calculations}, it suffices to run the simple algorithms of Lemma~\ref{lem:tightAlgo} and Theorem~\ref{thm:characterization} (with either $Q_1$ or $Q_2$) and check whether the resulting sequence is a valid $H$-recoloring sequence (if not, the assumption that some realizable walk exists was false).

Similarly, we can easily deduce the following purely graph-theoretic observation:
\begin{theorem}\label{lem:3col}
	Let $G$ be a graph with no cycles of length divisible by 3. Then $G$ is 3-colorable.
\end{theorem}
\begin{proof}
	The proof is by induction on the number of edges: let $\alpha$ be a 3-coloring of $G-e$, for an arbitrary $e=uv\in E(G)$.
	If $\alpha(u)\neq \alpha(v)$, then this is a 3-coloring of $G$.
	
	Otherwise, define $\beta(x)=\alpha(x) + 1 \mod 3$ (where the colors, or vertices of $K_3$, are $\{0,1,2\}$).
	This is another 3-coloring of $G-e$, obtained just by rotating $\alpha$, so homotopic to $\beta$.
	That is, we can choose $q\in V(G)$ arbitrarily and let $Q$ be the walk of length 2 from $\alpha(q)$ to $\beta(q)$: it has even length, it is easily checked to be topologically valid, and by the assumption that $G$ has no cycles of length divisible by 3, there are no $\alpha$-tight cycles.
	Hence $Q$ is realizable for $\alpha,\beta,q$.
	
	Therefore, there is a $H$-recoloring sequence from $\alpha$ to $\beta$.
	But then at some point $u$ or $v$ changes its color for the first time, so it becomes different from the color of $v$ or $u$, respectively, giving a 3-coloring of $G$.
\end{proof}

The statement already follows from a stronger theorem of Chen and Saito~\cite{ChenS94}, that graphs with no cycles of length divisible by 3 are in fact 2-degenerate (all their subgraphs have a vertex of degree $\leq 2$).
But at least in principle, this shows we can deduce the existence of homomorphism using reconfiguration.

If we only exclude cycles of length divisible by 3 as \emph{induced} subgraphs, it is an open problem whether such a graph is 3-colorable, but Bonamy et al.~\cite{BonamyCT14} recently showed that the chromatic number is bounded.
In~\cite{workshop}, Bonamy points out that 3-colorability would follow from the same proof as Theorem~\ref{lem:3col} if the following were true:
\begin{conjecture}
Every graph $G$ without induced cycles of length divisible by 3 has an edge $e$ such that $G-e$ still has no induced cycles of length divisible by 3.
\end{conjecture}

Curiously, the chromatic number of graphs with no induced cycles of length divisible by 3 is related to a very different relation between colorings and topology conjectured by Gil Kalai and Roy Meshulam, see~\cite{BonamyCT14}.
See also~\cite{BrewsterMMN16} and~\cite{SeymourS16} for further results on coloring graphs with few cycles of prescribed mod $k$ length.

\subsection*{Generalizations}
We note that none of the proofs in this paper used any structural properties of $H$.
If we consider \problem{$H$-Recoloring} for any graph $H$, but only allow recoloring a vertex if all of its neighbors have one common color (in other words, a reconfiguration step is allowed only when the homotopy class of the mapping does not change), the same results will follow.

An obvious question is how far can our results be extended to more general CSPs: to the asymmetric case, to multiple constraints, to hypergraphs (relations of arbitrary arity)?
Is there any connection with the tractable cases of generalized SAT reconfiguration problems?

Another question is whether the problems of graph homomorphism reconfiguration exhibit a dichotomy.
For which graphs $H$ is \problem{$H$-Recoloring} in \cclass{P} or \cclass{PSPACE}-complete?
For the hard side, it is known that \problem{$K_4$-Recoloring} is \cclass{PSPACE}-complete even for $G$ bipartite~\cite{BonsmaC09}.
This is equivalent to saying that \problem{$H$-Recoloring} is \cclass{PSPACE}-complete for $H$ the cube graph $K_4 \times K_2$,
which similarly implies that \problem{$H$-Recoloring} is \cclass{PSPACE}-complete for $H$ the 4-cycle $C_4$ with all loops added, for example.
An easy reduction (known as \emph{folding}, see~\cite{FieuxL12}) allows us to focus on so called \emph{stiff} graphs.
These statements are discussed in more detail in~\cite{Wro14thesis}.

Finally, it could be interesting to explore the implications of the square-free property for the whole Hom complex.

\subsection*{Acknowledgments}
The author would like to thank Amer E. Mouawad and Naomi Nishimura for helpful discussions and their hospitality, and anonymous referees for their many remarks.
Many thanks to Jarosław Błasiok for sharing his knowledge of algebraic topology, in a remarkably concise way.

\vspace*{-0.8\baselineskip}
\printbibliography

\end{document}